\documentstyle[epsf,sprocl,overcite,refsty_p]{article}

\def\be{\begin{equation}}
\def\ee{\end{equation}}
\def\bea{\begin{eqnarray}}
\def\eea{\end{eqnarray}}

\begin{document}

\titlepage

\vbox to 0pt{
\vspace*{-1.5cm}
\begin{flushright}
hep-ex/9703014\\
March, 1997
\end{flushright}
\vss}

\title{DEEP INELASTIC SCATTERING FROM UNPOLARISED TARGETS}

\author{ G.~K. MALLOT }

\address{Institut f\"ur Kernphysik,
Universit\"at Mainz,
Becherweg 45,\\
D-55099 Mainz, Germany}


\maketitle\abstracts{
In this paper the experimental status of unpolarised 
structure functions is reviewed. In particular the latest 
results from the NMC, E665, CCFR, and HERA experiments
are discussed.  Emphasis is put on the fixed-target experiments, 
which cover with high precision the $x$ region relevant for 
the present polarised DIS experiments.
}
\vfill
\begin{center}
Invited Talk at the \\
\it
International School of Nucleon Structure\\
1st Course: The Spin Structure of the Nucleon\\
\rm
Erice, Italy, 3--10 August 1995
\end{center}

\newpage
\setcounter{page}{1}
\title{DEEP INELASTIC SCATTERING FROM UNPOLARISED TARGETS}

\author{ G.~K. MALLOT }

\address{Institut f\"ur Kernphysik,
Universit\"at Mainz,
Becherweg 45,\\
D-55099 Mainz, Germany}

\maketitle\abstracts{
In this paper the experimental status of unpolarised 
structure functions is reviewed. In particular the latest 
results from the NMC, E665, CCFR, and HERA experiments
are discussed.  Emphasis is put on the fixed-target experiments, 
which cover with high precision the $x$ region relevant for 
the present polarised DIS experiments.
}

\section{Introduction}
Deep inelastic scattering (DIS) of charged and neutral leptons
has revealed most of what we know about the quark structure
of the nucleon and of nuclei. The interpretation of DIS data is
based on the factorisation of the hard scattering process from
the nonperturbative nucleon structure. The former is described
by Quantum Chromodynamics (QCD) while the latter is parametrised in
terms of structure functions. Thus the objectives of DIS experiments are
twofold. They test QCD and probe the nucleon's structure at the same 
time.
DIS experiments led to the discovery of partons, which later were 
identified with the hypothetical quarks, postulated earlier
to explain the hadron spectrum. 
It was also found that the charged partons are fermions with spin 
1/2 and  carry only about half of the nucleon's momentum. 
This suggests that gluons play an important r\^ole in the nucleon.
The same partition was recently predicted for the angular 
momentum~\cite{JiT96a} in the limit of infinite momentum transfer,
$Q^2\rightarrow\infty$.
With the advent of the high statistics experiments at large $Q^2$, the 
gluon distribution could be inferred from the $Q^2$ dependence of
the structure functions using the Gribov--Lipatov--Altarelli--Parisi
evolution equations (GLAP)~\cite{GrL72,Dok77,AlP77}.
Important input for the flavour decomposition of the quark content 
of the nucleon came from charged-current neutrino scattering.
It was shown that the nucleon contains three valence quarks and that the
mean-square charge of the up and down quarks is 5/18. 
From opposite-sign di-muon events the distribution of the strange
quarks was obtained.~\cite{CCFR95}

Although a consistent picture emerged and DIS developed to a precise tool,
its history is accompanied by surprises. 
First strong effects due to nuclear binding~\cite{EMC83a} were discovered by 
the European Muon Collaboration, the so-called EMC-Effect. 
Later the violation  of the Ellis--Jaffe 
sum rule,~\cite{ElJ74} discovered by the EMC~\cite{EMC88a,EMC89a} in polarised
muon-proton scattering,
questioned our understanding of the nucleon's spin structure. 
Then the violation~\cite{NMC91b,NMC94a} of the Gottfried sum rule~\cite{Got67}
found by the New Muon Collaboration (NMC) revealed that the light quark sea is 
not flavour symmetric.
Recently, a strong increase of $F_2$ at very small $x$ was found at 
HERA~\cite{H1_93a,ZEUS93a} and finally, the excess of events at high $x$ and 
$Q^2>15000~\mbox{GeV}^2$ also seen at HERA~\cite{H1_97a,ZEUS97a} might be a
trace of ``new physics''.

\begin{figure}[t]
   \begin{minipage}[t]{0.39\hsize}
      \begin{center}
         \vbox{
         \mbox{\epsfxsize=\hsize\epsfbox[10 90 590 360]{figures/dis.eps}}\\
         \vspace{2cm}
          }
         \end{center}
      \caption{The deep inelastic scattering process.}
      \label{fig:dis}
      \end{minipage}
   \hfill
   \begin{minipage}[t]{0.59\hsize}
      \begin{center}
         \mbox{\epsfxsize=\hsize\epsfbox[46 160 520 654]{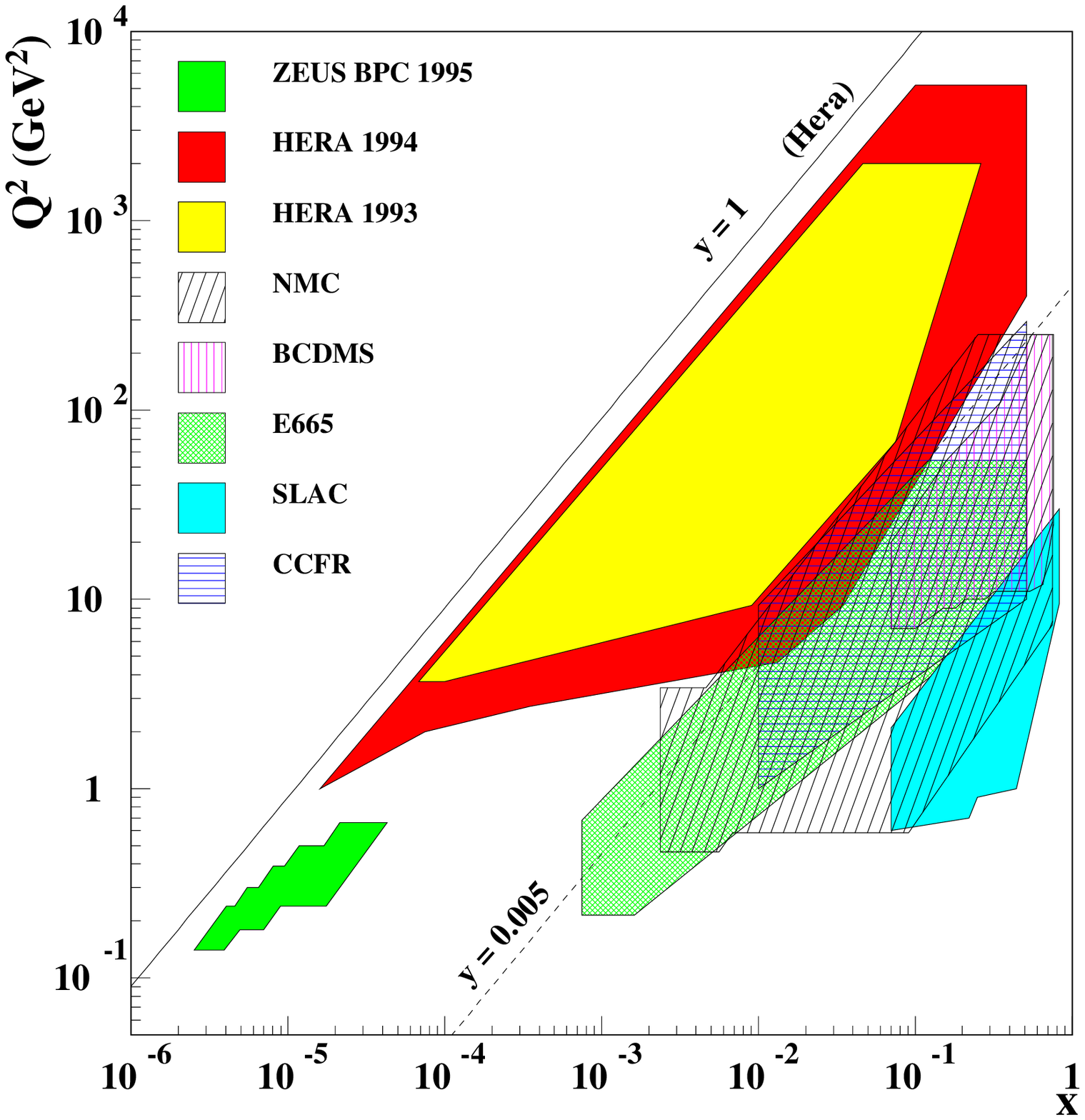}}
        \end{center}
      \caption{Kinematic range~\protect\cite{BaR96} of the HERA, NMC, BCDMS, 
       E665, SLAC, and CCFR experiments.}
      \label{fig:range}
      \end{minipage}
   \end{figure}

The DIS process is sketched in Fig.~\ref{fig:dis}. 
The kinematics in the laboratory frame for fixed-target experiments
is given by the incoming lepton energy, $E$, the energy transfer, $\nu=E-E'$,
the square of the virtual-photon mass, $q^2=-Q^2$, and the proton mass, $M$.
In Born approximation the inclusive, differential cross section in terms of the
scaling variables, $x=Q^2/2pq=Q^2/2M\nu$ and $y=pq/pk=\nu/E$, reads
\begin{equation}
\label{eq:cross}
\frac{{\rm d}^2\sigma^{iN}}{{\rm d}x{\rm d}y}=
a^{iN}  \left\{
        xy^2 F_1^{iN}+
        (1-y-\frac{xyM^2}{s-M^2})F_2^{iN}
        \pm(y-\frac{y^2}{2})xF_3^{iN}
        \right\},
\end{equation}
where $F(x,Q^2)$ are the dimensionless structure functions and
$s=Q^2/xy+M^2$ is the square of the lepton-nucleon c.m.\ energy.
The coupling constants and the boson propagators yield,
\be
a^{\ell\rm N}=4\pi\alpha^2(s-M^2)\frac{1}{Q^4}\hskip 3mm \mbox{and}\hskip 3mm
a^{\nu\rm  N}=\frac{G_F^2}{2\pi}(s-M^2)\frac{1}{(1+Q^2/M_W^2)^2},
\ee
for charged-lepton ($i=\ell$) and charged-current neutrino scattering 
($i=\nu,\overline{\nu}$), respectively. 
The parity-violating structure function $F_3$ enters with positive (negative)
sign for (anti)neutrino scattering and vanishes for electron and muon
scattering. Neutral-current neutrino scattering and charged-current 
electron (muon) scattering~\cite{H1_96b,ZEUS96c} are not discussed in this paper.

In the Quark-Parton-Model (QPM), which is motivated in the
Bjorken limit, $Q^2, \nu\rightarrow \infty$ at fixed $x$, the scaling
variable $x$ represents the
fraction of the nucleon's longitudinal momentum carried by the struck quark.
The structure functions become functions of $x$ only and acquire a very
intuitive interpretation in terms of quark and antiquark distribution 
functions, 
$q_f(x)$ and $\overline{q}_f(x)$ with $q_f= u, d, s, c, b, t$.
One finds the Callan--Gross relation~\cite{CaG69}, $2xF_1(x)=F_2(x)$, and
\begin{eqnarray}
\label{eq:sf_qpm}
F_2^{\ell\rm N}(x) & = & x \sum_f e_f^2 \{ q_f(x)+\overline{q}_f(x)\},\nonumber\\
F_2^{\nu\rm N}(x) = F_2^{\bar\nu\rm N}(x)
                & = &x \sum_f \hskip 4.1mm \{ q_f(x)+\overline{q}_f(x)\},\\
\frac{1}{2}x\left[F_3^{\nu\rm N}(x)+F_3^{\bar\nu\rm N}(x)\right] 
                & = &x \sum_f \hskip 4.1mm \{ q_f(x)-\overline{q}_f(x)\},\nonumber
\end{eqnarray}
where $e_f$ denotes the electric charge of a quark with flavour $f$.
In Eq.~\ref{eq:sf_qpm} neutrino scattering from isoscalar targets is assumed 
and in addition $s(x)=\overline{s}(x)$ and
$c(x)=\overline{c}(x)$ is used in the expressions for $F_2^{\nu\rm N}$ and 
$F_2^{\bar\nu\rm N}$.
In the QCD-improved QPM a logarithmic $Q^2$ dependence of the structure functions
is generated by gluon bremsstrahlung. 
This $Q^2$ dependence can be calculated in QCD and is described by the GLAP 
equations. It is one of the cleanest tools to determine the gluon distribution 
function of the nucleon, $g(x,Q^2)$, and  the strong coupling constant 
$\alpha_s$.

In this paper emphasis is put on the results relevant to the analysis of the
polarised DIS data, i.e.\ the region $0.001<x$ and the main part of the paper
is dedicated to fixed-target experiments. However, the most important
results from HERA are also reviewed.
The results from the Zeus experiment at HERA are discussed in more detail
in a separate contribution~\cite{Zeuner} to this Workshop.
Due to the lack of space a discussion of nuclear effects in structure 
functions has been omitted.

\section{The Experiments}
The series of {\em unpolarised\,} electron DIS experiments at SLAC began in the late 
1960's~\cite{TKF91} and lasted till 1985~\cite{WhR92} and beyond.
The scattered electrons were
detected by 1.6, 8, and 20~GeV/$c$ small-aperture magnetic spectrometers. 
The maximum incident electron energy was 20~GeV covering the kinematic range
$x\ge 0.07$.

In the 1980's the high-intensity 280~GeV CERN muon beam served simultaneously the 
experiment of the BCDMS and that of the
European Muon Collaboration (EMC).~\cite{SaS88} 
The momentum of the incident muon was
measured by a dedicated common magnetic spectrometer.
The EMC apparatus comprised an extended target area with an about 5~m long
liquid hydrogen or deuterium target, a large-aperture spectrometer magnet
and a muon-identification stage downstream of a hadron absorber. 
Subsequently upgraded versions of this spectrometer~\cite{EMC81a} were used 
by the New Muon (NMC) and by the Spin Muon Collaboration (SMC). 
The NMC experiments were optimised for the determination of cross-section ratios.
A group of two targets along the beam axis was frequently exchanged with
a second group of targets, in which the order of the target materials along
the beam was inverted. This yielded very precise results for
$F_2^{\rm n}/F_2^{\rm p}$~\cite{NMC96b} and for nuclear effects in the 
structure functions.~\cite{NMC95a,NMC96d}
The BCDMS experiment consisted of a series of segmented toroids 
interspersed with \mbox{MWPCs} and trigger hodoscopes.
The central bore contained the in total 40~m long targets.
The muons were bent by the magnetic field in the iron toroids towards
the axis of the spectrometer.
This setup resulted in a good acceptance for large scattering angles,
i.e.\ for large values of $x$ and $Q^2$.
The to date last unpolarised muon DIS experiment was performed 
by the E665 Collaboration~\cite{E665_96a} using the
470~GeV muon beam at FNAL.
The principle of the spectrometer is similar to that of the EMC. 
However, it involved an additional spectrometer magnet close to the target.

In the early 1980's also charged-current neutrino--nucleon scattering
experiments started at CERN with the CDHSW experiment.~\cite{CDHSW91a} 
The most precise data today come from the CCFR Collaboration at FNAL, 
which took data in the late 1980's.
Neutrinos and antineutrinos of 30--600~GeV from kaon and pion decays 
impinged on a 690~ton iron target and the emerging muons were detected
in a 420 ton toroid system. 
Apart from $F_2^{\nu\rm N}$ and $F_3^{\nu\rm N}$ results for the
Gross-Llewellyn Smith sum rule~\cite{CCFR93a} and the strange quark content
of the nucleon~\cite{CCFR95} were obtained.

Finally, in the 1990's the HERA e-p collider at DESY with its H1 and 
ZEUS~\cite{Zeuner} experiments opened up a completely new kinematic domain.
The 820~GeV proton and the 27~GeV electron beam provide a centre-of-mass
energy of $\sqrt{s}\simeq300~$GeV with which $x$ values as low as
$x=10^{-4}$ can be reached at a momentum transfer of 
$Q^2\simeq5~\mbox{GeV}^2$ and for $x>0.1$ values of 
$Q^2\ge 5000~\mbox{GeV}^2$ can be accessed.
The kinematic range of the individual experiments is shown in 
Fig.~\ref{fig:range}.

\begin{figure}[t]
\begin{center}
\mbox{\epsfxsize=0.7\hsize\epsfbox[60 200 550 860]{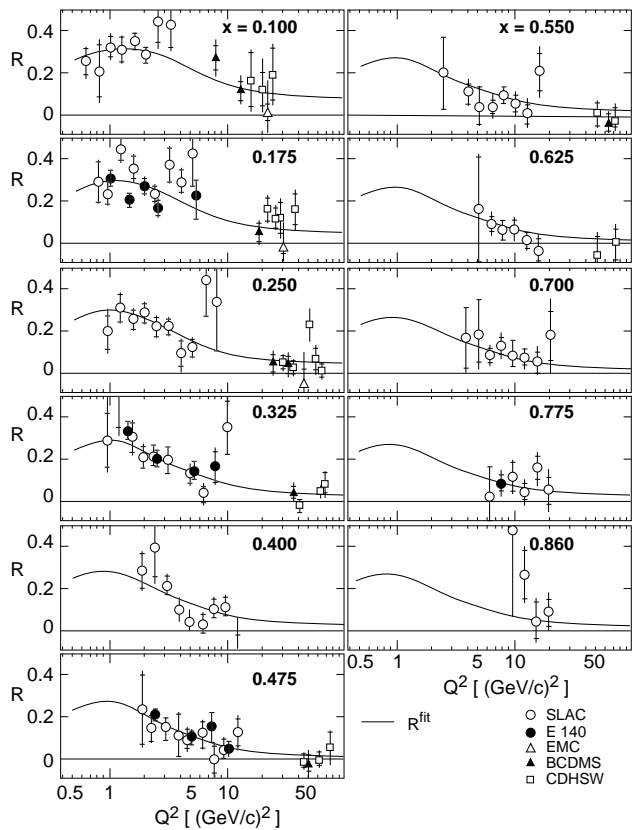}}
\end{center}
\caption{
SLAC analysis of $R(x,Q^2)$ data.~\protect\cite{WhR90}
The solid curve is a parametrisation of the data, which is widely used (R1990).}
\label{fig:R1990}
\end{figure}

\section{The Longitudinal-to-Transverse Cross-Section Ratio}
Both, longitudinally and transversely polarised photons contribute to the
differential cross section of Eq.~\ref{eq:cross}. 
The ratio of the corresponding cross sections, $\sigma_L$ and $\sigma_T$,
is given by
\be
\label{eq:R}
R=\frac{\sigma_{L}}{\sigma_{T}} = \frac{F_2(1+Q^2/\nu^2)}{2xF_1}-1 = 
                                          \frac{F_L}{2xF_1}.
\ee
It can determined from  measurements at the same ($x$, $Q^2$) point at
different values of $y$, i.e.\ different incident lepton energies.
Transverse quark momenta introduced by gluon bremsstrahlung allow also the 
absorption of longitudinal virtual photons. 
Due to the absence of transverse momenta, $F_L$ vanishes in the na\"\i ve QPM.
In perturbative QCD $F_L$ and thus $R$ ($R_{\rm QCD}$) can be 
calculated~\cite{AlM78} from $F_2$ and from the gluon distribution function, $g$,
\be
\label{eq:FL}
F_L(x,Q^2) = \frac{\alpha_s}{\pi} x^2\left\{
\frac{4}{3} \int_x^1 F_2(y,Q^2) \frac{{\rm d}y}{y^3} +
2 c        \int_x^1 g(y,Q^2)
           \left(1-\frac{x}{y}\right) 
           \frac{{\rm d}y}{y^2}\right\},
\ee
where $c=\sum e_f^2$ for electron and muon scattering and $c=n_f$, the 
number of active flavours, for neutrino scattering.
Note that $F_L$ is proportional to the strong coupling constant 
$\alpha_s(Q^2)$.
At small values of $x$, where $g(x,Q^2)$ rises sharply, an increase of 
$F_L$ is expected from Eq.~\ref{eq:FL}.

\begin{figure}[t]
\mbox{\epsfxsize=0.45\hsize\epsfbox[17 211 520 671]{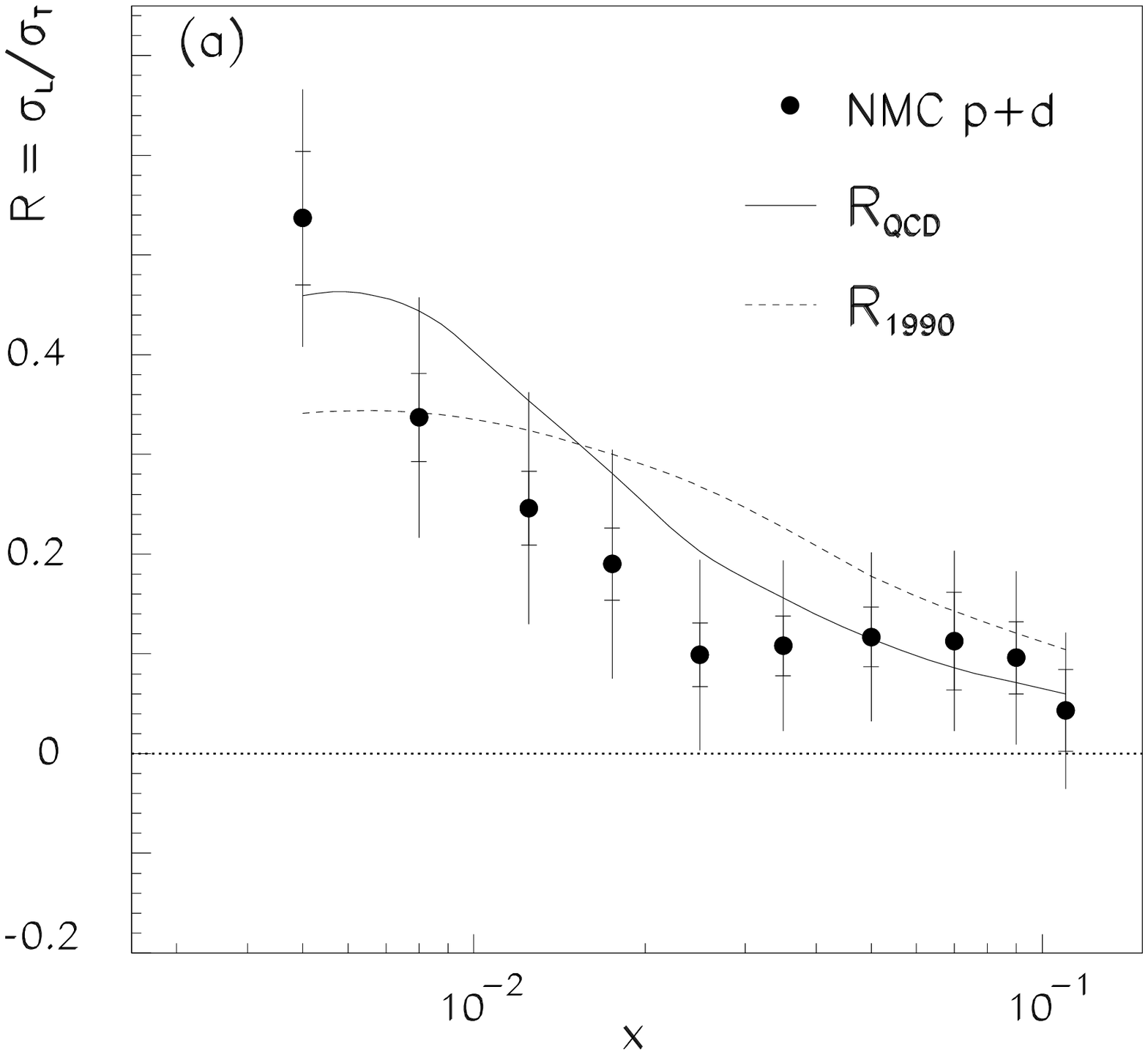}}
\hfill
\mbox{\epsfxsize=0.45\hsize\epsfbox[17 211 520 671]{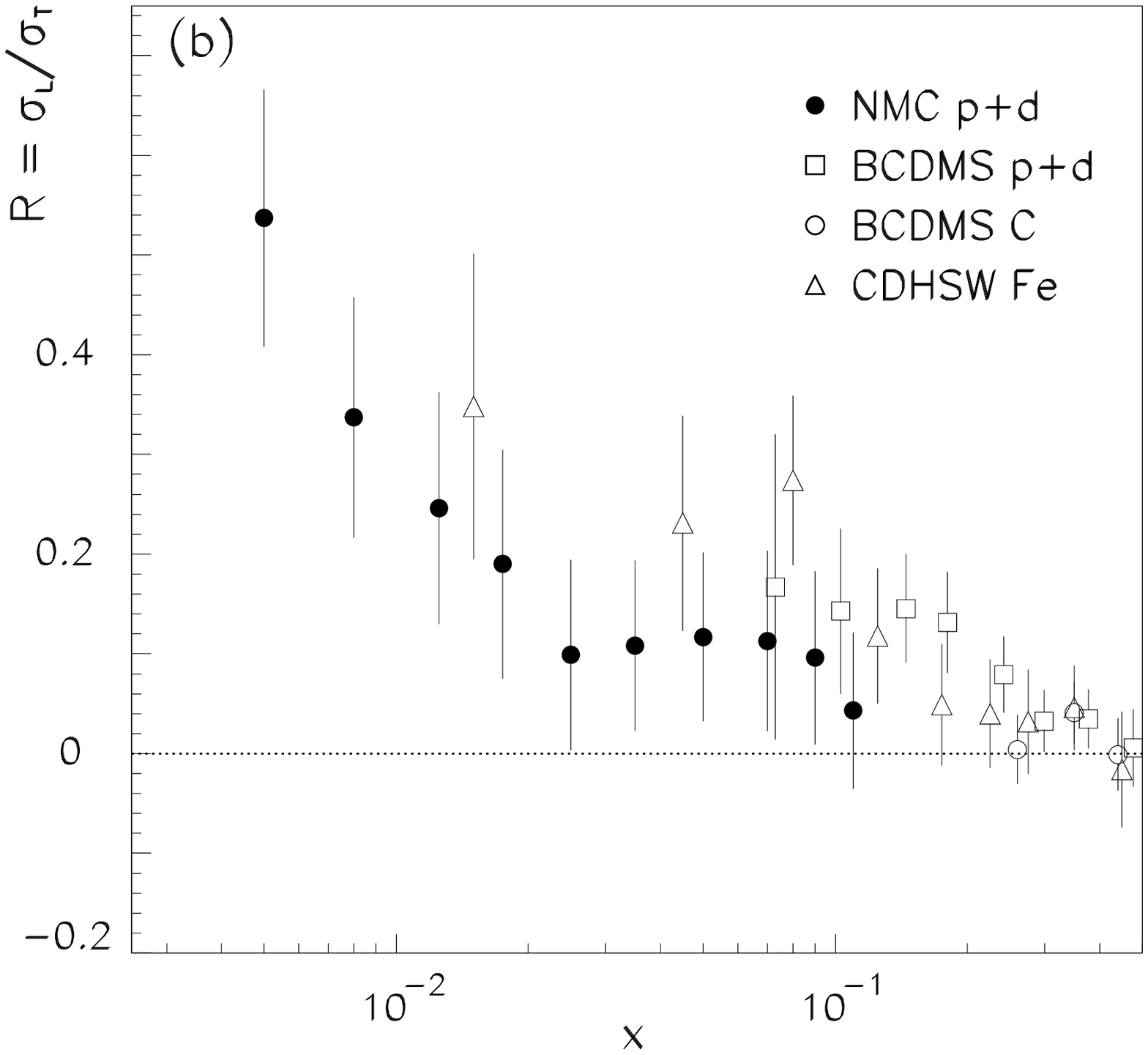}}
\caption{
(a) NMC measurement~\protect\cite{NMC96a} of $R$ as function of $x$ together 
with the R1990 (solid line) parametrisation~\protect\cite{WhR90} and 
$R_{\rm QCD}$ (dashed line). 
(b) Comparison with several other experimental results.}
\label{fig:NMC_R}
\end{figure}

The result of a comprehensive global analysis~\cite{WhR90} of the
the SLAC electron-proton and deuteron data on $R(x,Q^2)$ and some high $Q^2$ 
muon and neutrino data from CERN resulted in a phenomenological
parametrisation for $R$, often referred to as R1990, which is shown
together with the data in Fig.~\ref{fig:R1990}.
The kinematic region covered is
$0.1\le x \le 0.9$ and $0.6<Q^2<20$~GeV$^2$ for the SLAC data
and up to about 80~GeV$^2$ for the muon and neutrino data.
The ratio $R$ is rather well determined for intermediate $x$
and large $Q^2$, where it is small. However, the behaviour in the region
$x<0.1$ is uncertain. 
Recently, new data from the NMC~\cite{NMC96a} in the small-$x$ region,
$0.002<x<0.12$, became available (Fig.~\ref{fig:NMC_R}).
The data agree well with the rise expected from perturbative QCD 
and
-- in the region of overlap -- with the previous measurements. Also
the preliminary neutrino data for $R$ from
the CCFR Collaboration~\cite{CCFR96a} for $0.01<x<0.6$ and 
$Q^2>4~\mbox{GeV}^2$ agree well with $R_{\rm QCD}$ (not shown here).
The comparison of the neutrino $R$ data taken with an iron target and the
charged-lepton data shows no evidence for a
dependence of $R$ on the target material. Similar conclusions were
previously reached from muon~\cite{NMC92f,NMC96b} and electron 
experiments~\cite{E140_88,E140_94,E140X96} using a series of nuclear 
targets.

A measurement of $F_L$  at very small $x$  is presently being
considered at HERA using lower beam energies. First results were obtained
from the 1994 H1 data~\cite{H1_96y} using a different approach. 
Equation~\ref{eq:cross} can be rewritten 
in terms of $F_2$ and $F_L$ yielding a differential
cross section proportional to $(2-2y+y^2)F_2-y^2F_L$. 
Thus at small $y$ the cross section is dominated by $F_2$, while
$F_2$ and $F_L$ contribute with similar weight at large $y$. 
The longitudinal structure function $F_L$ was determined from the data with
$0.6\le y\le0.78$ by subtracting the $F_2$ contribution, which was obtained
from data with $y\le 0.35$. Since both data sets were taken using same
beam energies, the $F_2$ data had to be evolved from the measured $Q^2$ to 
the value corresponding to the higher value of $y$
using next-to-leading order GLAP evolution. In this sense $F_L(x,Q^2)$ was 
not directly measured but inferred from data taken at different $(x,Q^2)$ points.
The result, $F_L=0.52\pm0.25$ at $x=0.00024$ and 
$Q^2=15.4~\mbox{GeV}^2$,  corresponds to $R\simeq0.50$ and agrees well with
expectations.

\begin{figure}[p]
\mbox{\epsfxsize=0.45\hsize\epsfbox[53 100 543 730]{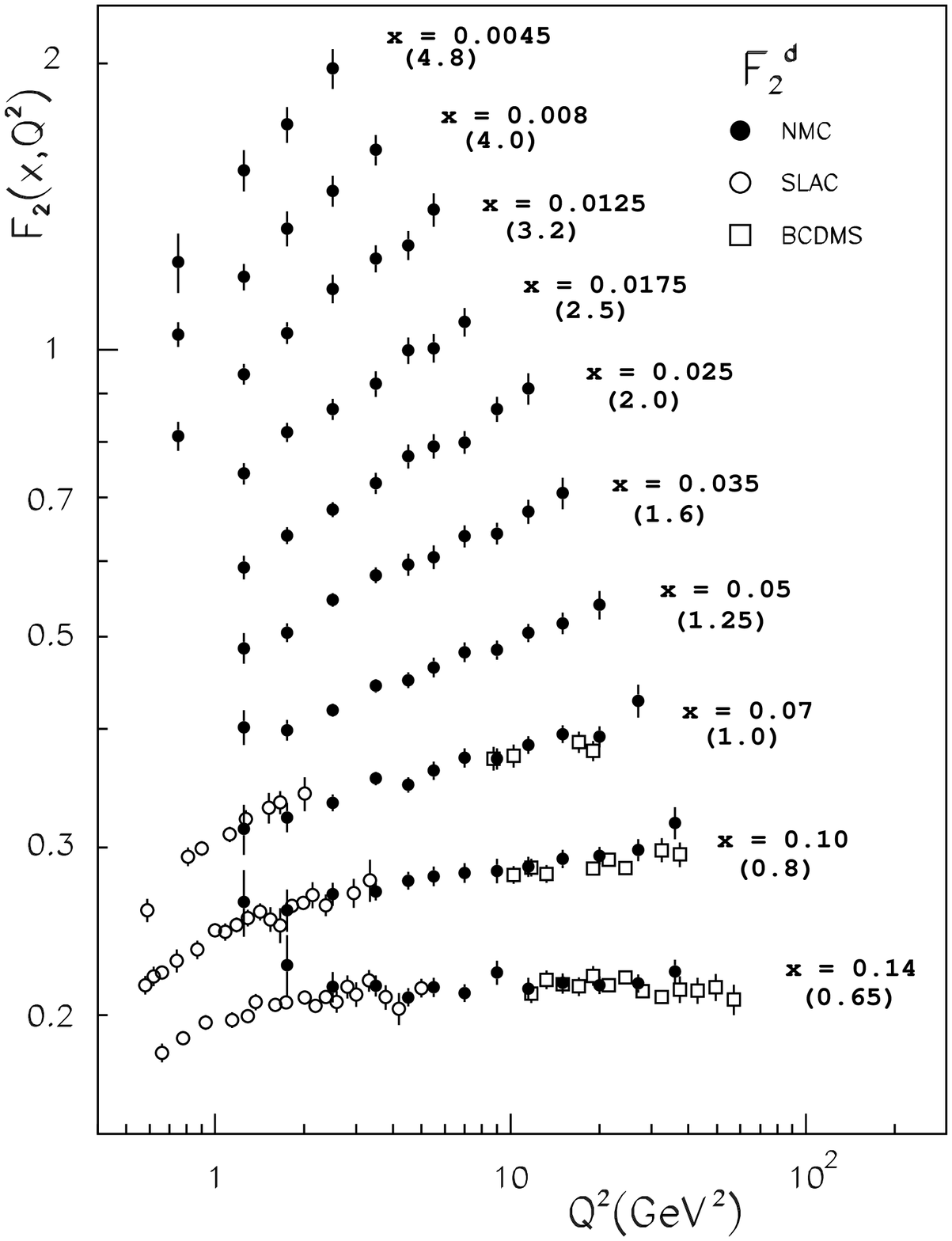}}
\hfill
\mbox{\epsfxsize=0.45\hsize\epsfbox[53 100 543 730]{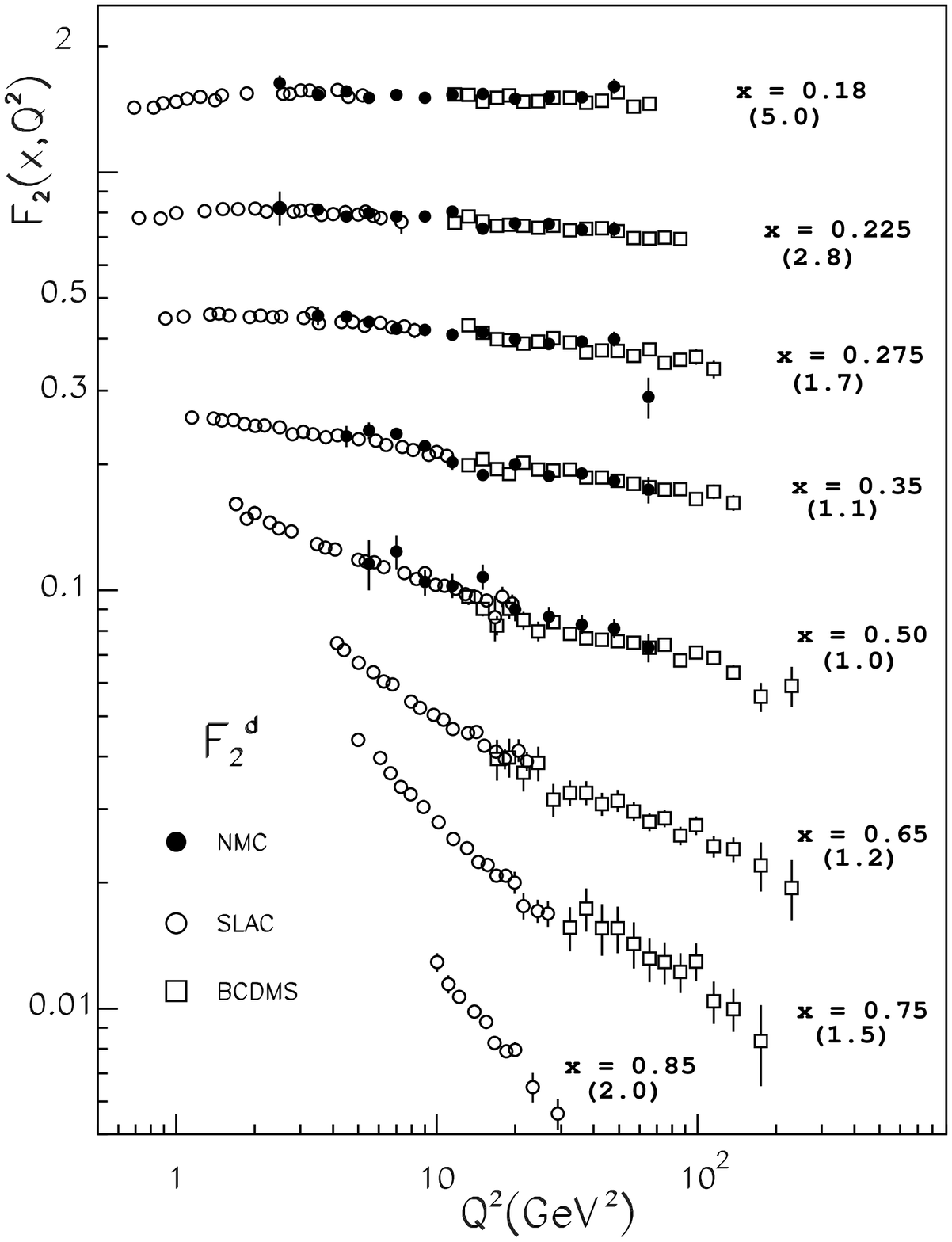}}
\caption{The structure function $F_2$ for the deuteron~\protect\cite{NMC96a} 
from the SLAC, BCDMS, and NMC experiments. 
The data were multiplied with the numbers in brackets for clarity of the plot.}
\label{fig:F2dall}

\bigskip
\begin{center}
\mbox{\epsfxsize=0.90\hsize\epsfbox[54 260 565 599]{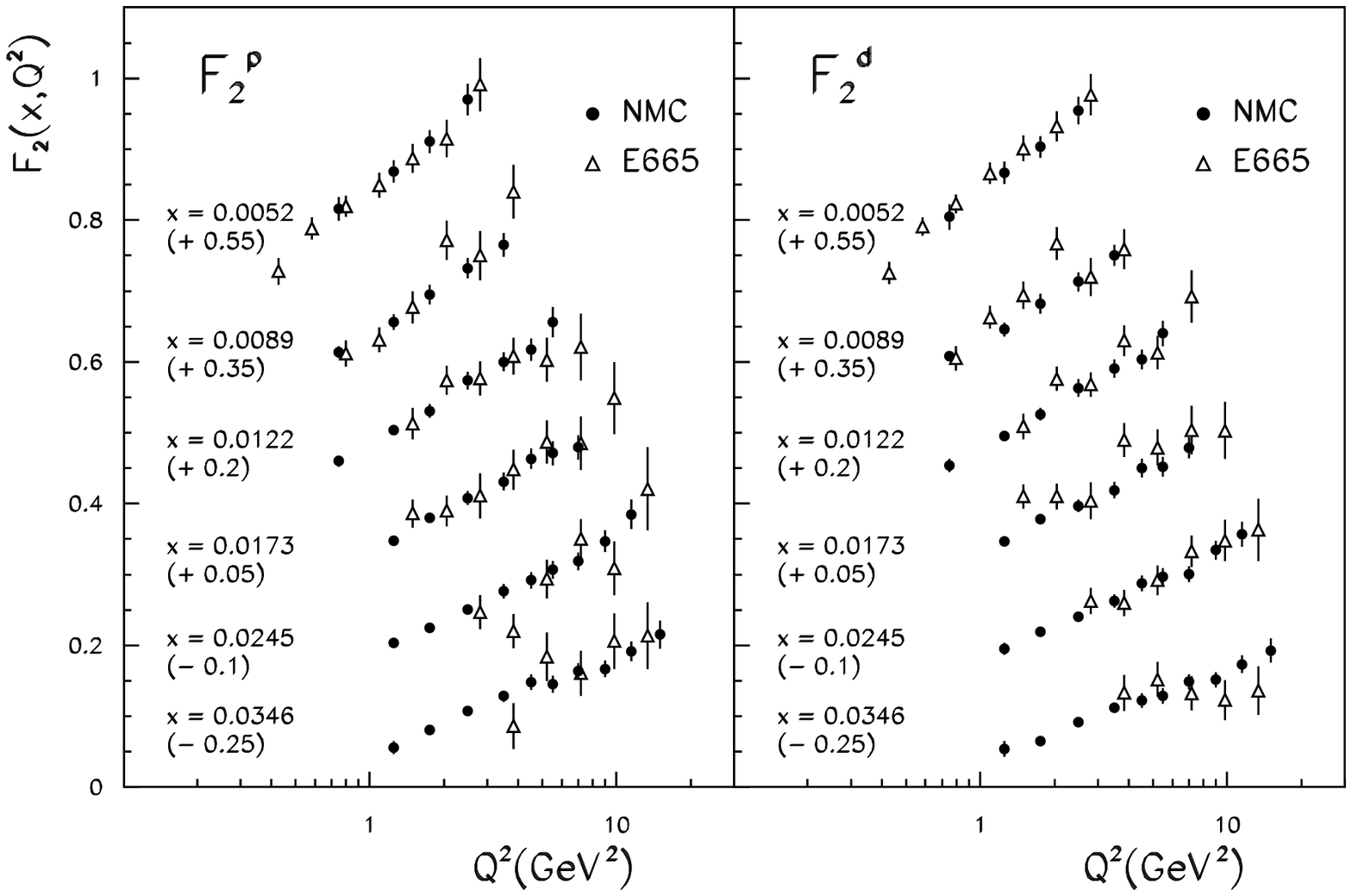}}
\end{center}
\caption{Comparison of the NMC~\protect\cite{NMC96a} and E665~\protect\cite{E665_96a} 
$F_2$ data for the proton (left) and the deuteron (right) in the range $0.004<x<0.04$. 
The values in the brackets were added to the data for clarity of the plot.}
\label{fig:F2_e665}
\end{figure}

\section{Structure Function Data}
\subsection{Muon and Electron Fixed-Target  $F_2(x,Q^2)$ Data}
\begin{figure}[t]
\begin{center}
\mbox{\epsfxsize=0.7\hsize\epsfbox[56 255 564 606]{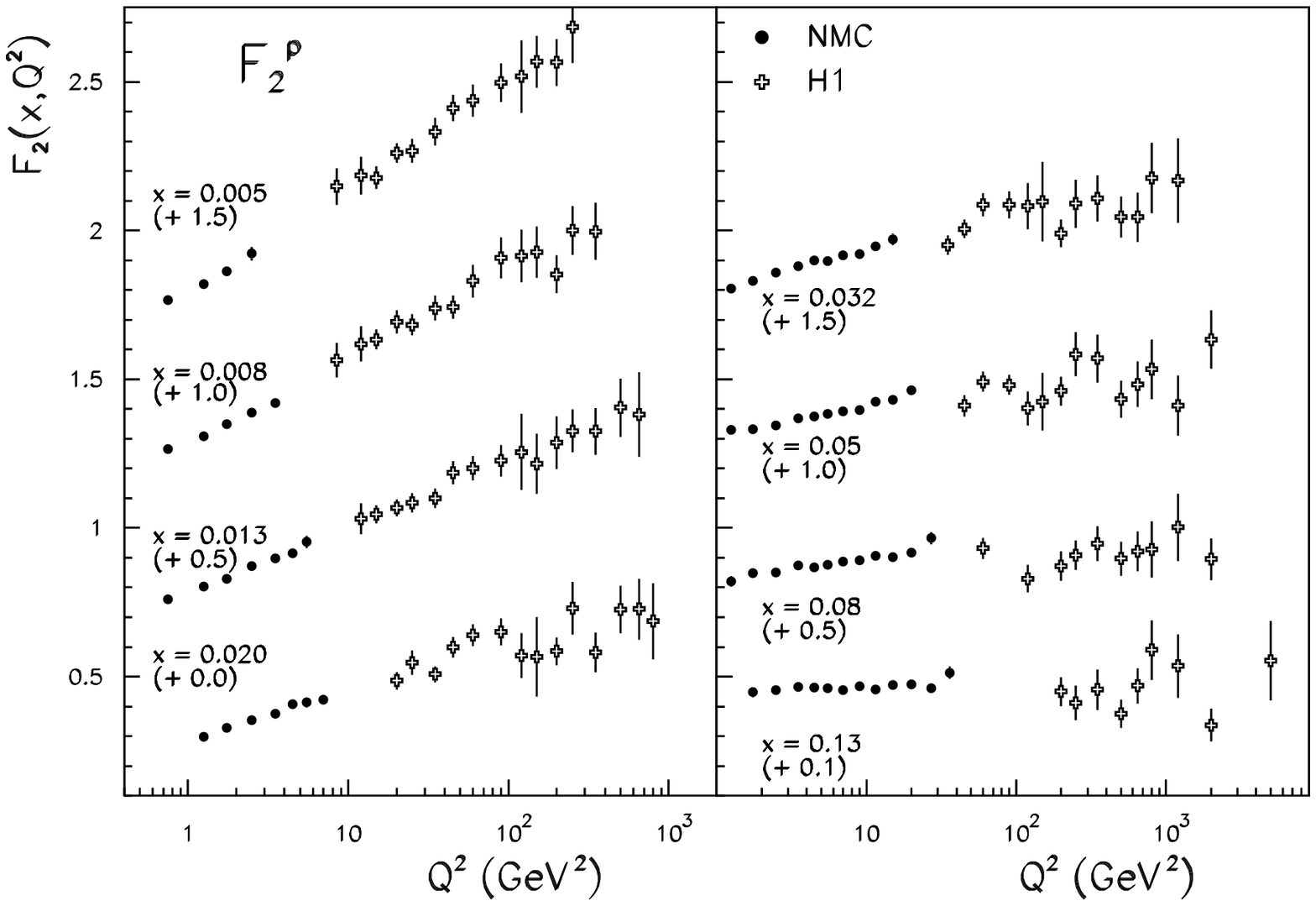}}\\
\mbox{\epsfxsize=0.7\hsize\epsfbox[56 255 564 606]{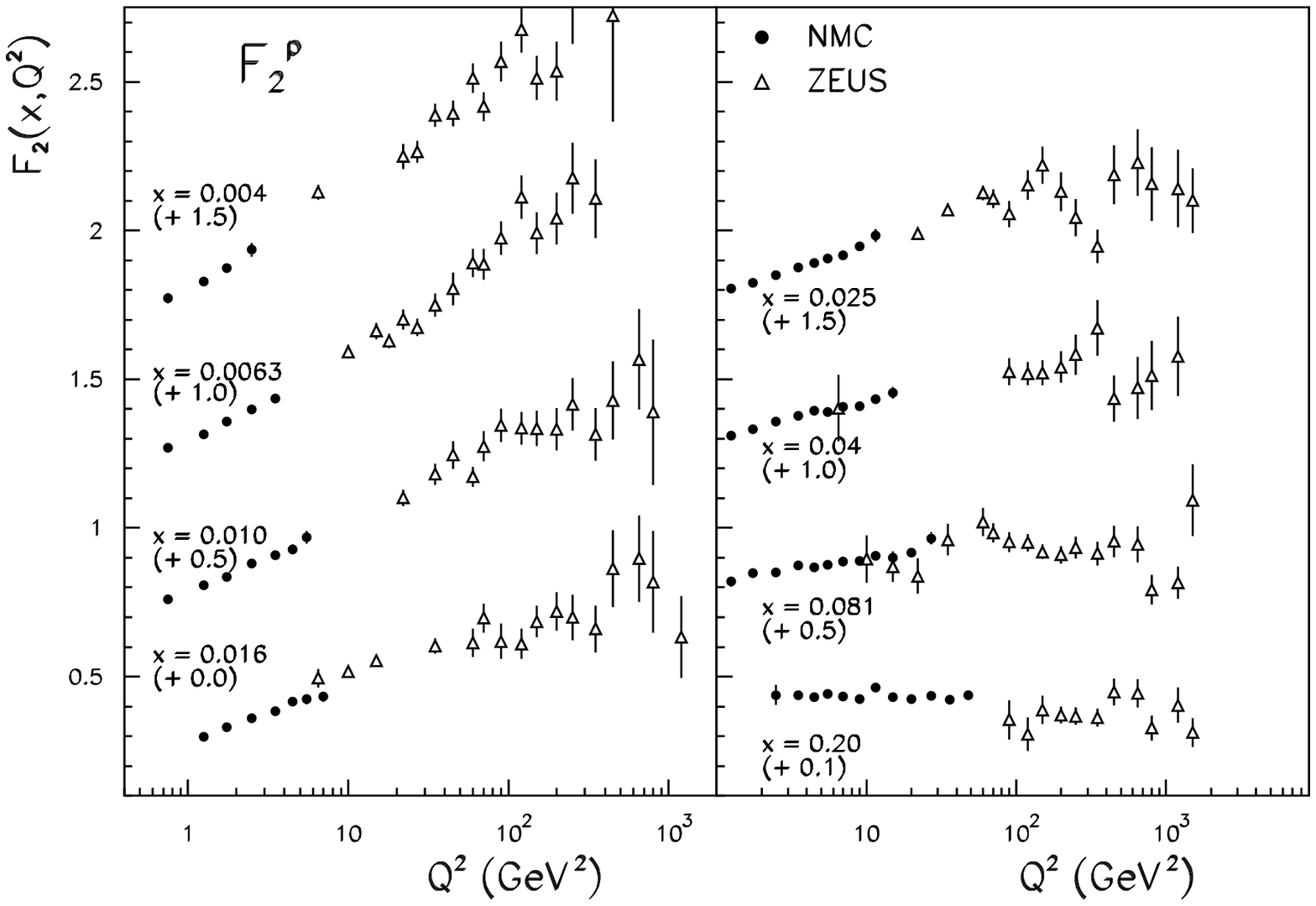}}
\end{center}
\caption{
Comparison~\protect\cite{NMC96a} of the NMC $F_2^{\rm p}$ data with the 
present HERA data from H1~\protect\cite{H1_96a} (top) and 
ZEUS~\protect\cite{ZEUS96a} (bottom) in the range $0.004<x<0.2$.}
\label{fig:NMC_HERA}
\end{figure}

The present status of the deuteron $F_2$ measurements from charged-lepton 
fixed-target experiments  is summarised  in Figs.~\ref{fig:F2dall}
and \ref{fig:F2_e665}. 
The proton data are in accuracy and kinematic coverage similar to the
deuteron results.
The data from SLAC,~\cite{WhR92} the NMC,~\cite{NMC96a} and the 
BCDMS~\cite{BCDMS89a,BCDMS90a} collaboration are in excellent
agreement.
They exhibit strong scaling violations, which are positive at small $x$
and negative at large $x$.
The EMC $F_2$ data~\cite{EMC85a,EMC87d} deviate systematically~\cite{MiS91} from
the BCDMS and NMC data at small $x$.
This discrepancy remained also after a re-analysis~\cite{EMC91y} 
of the EMC data, which therefore were not included in this 
compilation.
At large $x$ and $Q^2$ the data are dominated by the results
from the BCDMS while the small $x$ and $Q^2$ 
region is the domain of the SLAC data.
The final analysis of the NMC $F_2$ data comprises 
all the data taken with 90, 120, 200, and 280~GeV muon beams.
The NMC data provide the link between the SLAC and BCDMS data
and extend the measured $x$ region by more than an order of magnitude
from $x=0.07$ down to $x=0.0045$.
A useful phenomenological parametrisation to the NMC, BCDMS, and SLAC
data is given in Ref.~[\citen{NMC95c}].
A comparison of the NMC data with the E665 data~\cite{E665_96a} for the proton
and the deuteron in the range $0.004<x<0.04$ is shown in 
Fig.~\ref{fig:F2_e665}. 
The E665 data extent over five more small-$x$ bins not shown here down to
$x=0.0008$ where the average $Q^2$ is about 0.4~GeV$^2$. 
A comparison of the HERA data from H1~\cite{H1_96a} and Zeus~\cite{ZEUS96a}
to the NMC data, which are considerably more accurate,
is shown in Fig.~\ref{fig:NMC_HERA}.
The HERA data are in fair agreement with the  extrapolation of the NMC data 
to larger values of $Q^2$.

\begin{figure}[t]
\begin{center}
\mbox{\epsfxsize=0.7\hsize\epsfbox[0 150 452 681]{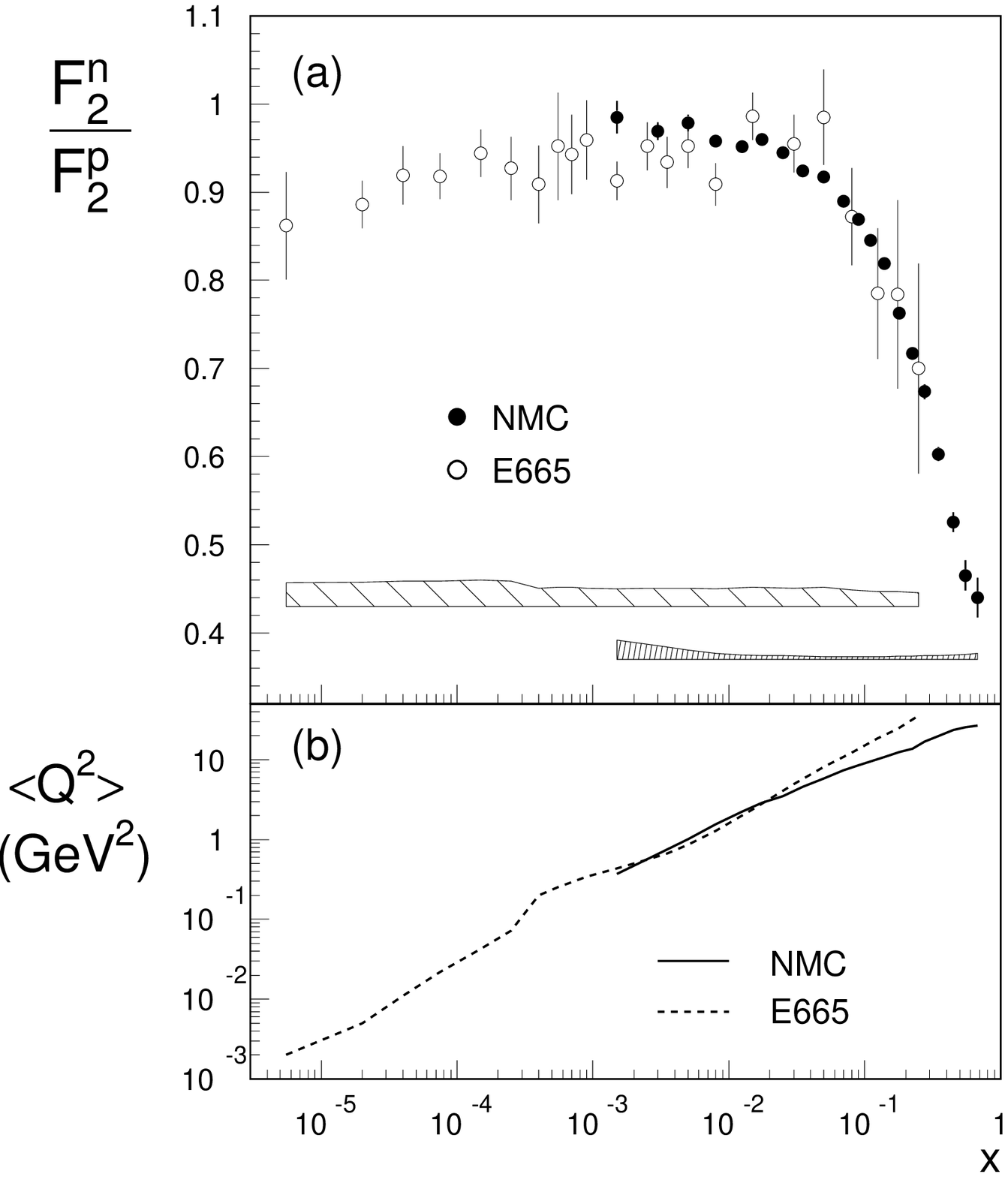}}
\end{center}
\caption{
(a) The neutron-to-proton structure function ratio, $F_2^{\rm n}/F_2^{\rm p}$,
as a function of $x$ from the NMC~\protect\cite{NMC96b} and 
E665.~\protect\cite{E665_95a} The shaded bands indicate the size of the systematic
uncertainties.
(b) Average $Q^2$ of the NMC and E665 data as a function of $x$.}
\label{fig:f2nf2p}
\end{figure}

\subsection{The ratio $F_2^{\rm n}/F_2^{\rm p}$ and the Gottfried sum rule}
Recent measurements of the
structure function ratio $F_2^{\rm n}/F_2^{\rm p}$ come from the
NMC~\cite{NMC96b} and the E665 Collaboration~\cite{E665_95a} 
(Fig.~\ref{fig:f2nf2p}).
The neutron-to-proton structure function ratio is obtained 
from experiments with liquid hydrogen and deuterium targets. Neglecting 
nuclear effects in the deuteron the ratio is given by
\be
\frac{F_2^{\rm n}}{F_2^{\rm p}} = 2\frac{F_2^{\rm d}}{F_2^{\rm p}}-1=
2\frac{\sigma^{\rm d}}{\sigma^{\rm p}}-1.
\label{eq:f2nf2p}
\ee
The latter step in Eq.~\ref{eq:f2nf2p} is justified by several 
measurements~\cite{WhR90,E140X96,NMC96b} finding the
difference $R^{\rm d}-R^{\rm p}$ to be compatible with
zero in the range $0.003<x<0.8$. This puts via Eq.~\ref{eq:FL} a 
limit~\cite{NMC96b} on a possible difference of the gluon distributions 
in the proton and the deuteron.
The ratio $F_2^{\rm n}/F_2^{\rm p}$ is found to be largely independent
of $Q^2$.
Small negative $Q^2$ slopes, 
${\rm d}(F_2^{\rm n}/F_2^{\rm p})/{\rm d}\ln Q^2$,
were observed~\cite{NMC96b} in the range $0.1<x<0.5$ in agreement with 
predictions from  perturbative QCD.
The ratio $F_2^{\rm n}/F_2^{\rm p}$ approaches unity for $x\simeq0.001$
as expected in this region where the sea quarks dominate.
The small deviation from unity is in the order of 0.02, which is a typical 
value expected for shadowing corrections in the deuteron.~\cite{BaK94,MeT93}
At very small $x$ the E665 data indicate a drop
below 0.9. Note however, that $Q^2$ is as small as 30~MeV$^2$ for 
$x=6\times10^{-6}$, what makes the data difficult to interpret.

The NMC used the ratio $F_2^{\rm n}/F_2^{\rm p}$ together with 
$F_2^{\rm d}$ to test the Gottfried sum rule~\cite{Got67}
\be
S_G=\int_0^1 \frac{F_2^{\rm p}-F_2^{\rm n}}{x}\,{\rm d}x
   =\frac{1}{3}+\frac{2}{3}\int_0^1(\bar u - \bar d ) \, {\rm d}x.
\label{eq:GSR}
\ee
If the last term vanishes the Gottfried sum rule takes its original
form, $S_G=1/3$. 
The NMC finds~\cite{NMC94a} $S_G=0.235\pm0.026$ at $Q^2=4~\mbox{GeV}^2$ 
significantly below 1/3. From Eq.~\ref{eq:GSR}
it is obvious that this result can be interpreted as a flavour-asymmetric sea
yielding $\int_0^1 (\bar u-\bar d)\,{\rm d}x=-0.165\pm0.059$. This explanation
is supported by the CERN experiment NA51, in which the cross-section asymmetry
$(\sigma^{\rm pp}- \sigma^{\rm pn})/(\sigma^{\rm pp}+\sigma^{\rm pn})$
for Drell-Yan production of muon pairs was measured using proton and deuteron 
targets.
From the asymmetry a value of $\bar u(x)/\bar d(x)=0.51\pm0.06$ at $x=0.18$ was 
inferred.~\cite{NA51_94} Both, the NMC and NA51 experiment, find a larger down than up-quark
component in the proton's quark sea.

\begin{figure}[t]
\begin{center}
\mbox{\epsfxsize=0.7\hsize\epsfbox[68 69 563 761]{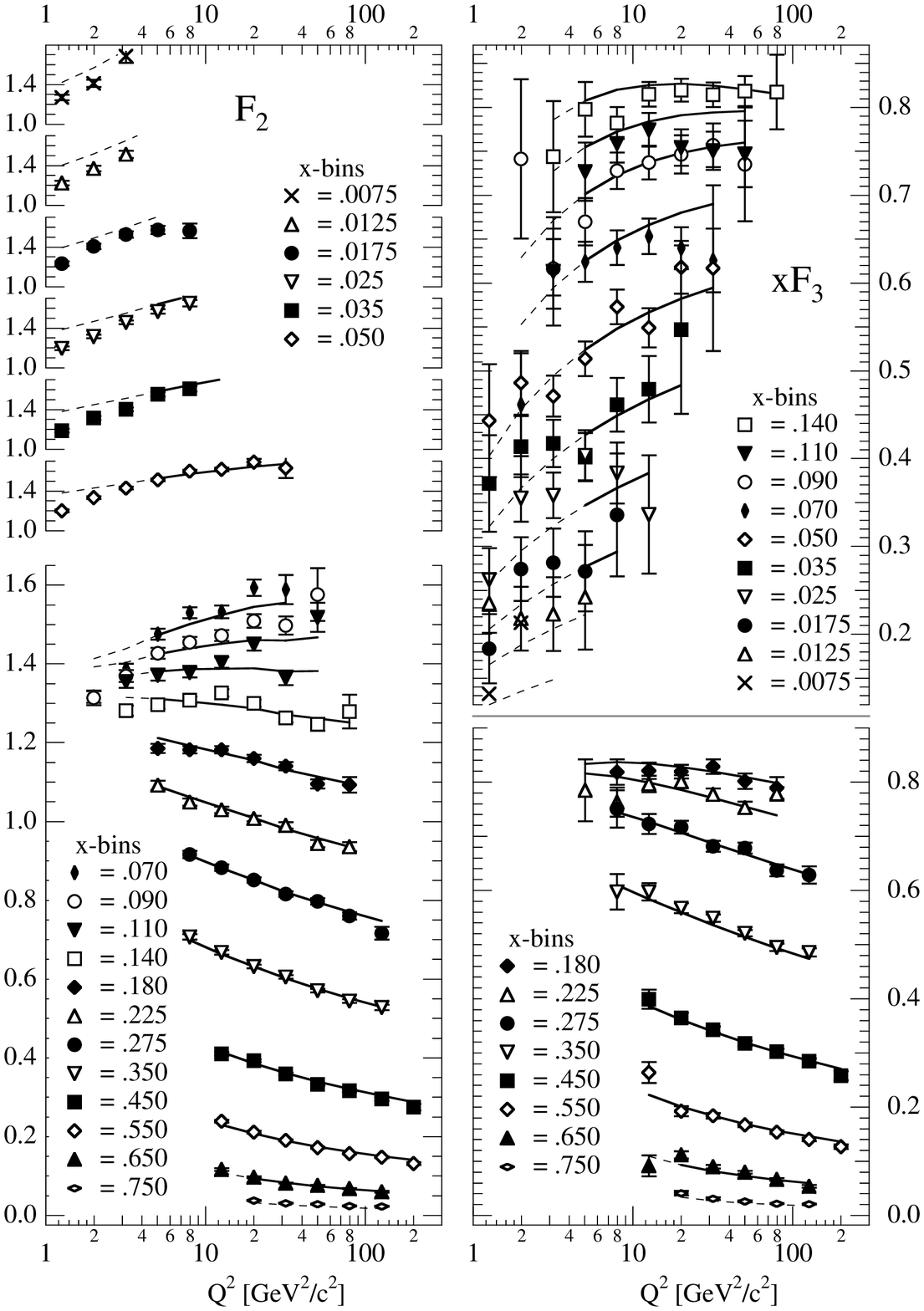}}
\end{center}
\caption{
The CCFR $F_2$ and $xF_3$ structure functions as a function of $Q^2$ for 
different values of $x$.~\protect\cite{CCFR97a}
Also shown is a QCD fit to the data (solid line) and its extrapolation to lower $Q^2$
(dashed line).}
\label{fig:ccfr_1}
\end{figure}

%
%
\begin{figure}
\begin{center}
\mbox{\epsfxsize=0.67\hsize\epsfbox[32 361 556 757]{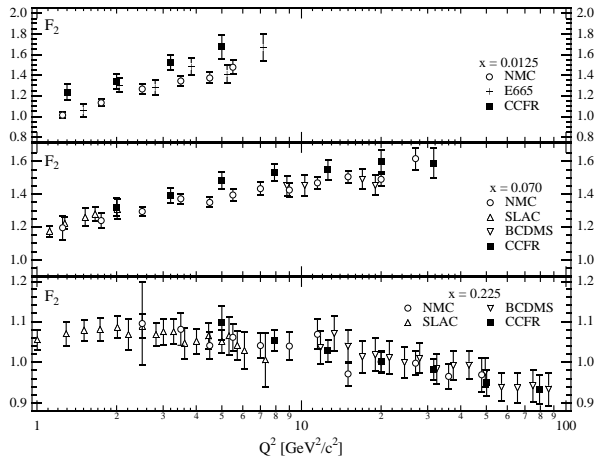}}
\end{center}
\caption{
Comparison of the $F_2$ data from neutrino and charged-lepton scattering
for the bins $x=0.0125$, 0.070, and 0.225.~\protect\cite{CCFR97a}}
\label{fig:ccfr_2}
\end{figure}

\subsection{Data from Neutrino Experiments}
The nucleon's structure functions, $F_2^{\rm N}=(F_2^{\rm p}+F_2^{\rm n})/2$,
measured in charged-lepton and neutrino  experiments are related by 
(see Eq.~\ref{eq:sf_qpm})
\be
\frac{F_2^{\ell\rm N}}{F_2^{\nu\rm N}}=
\langle e^2\rangle
   \left\{1-\frac{3}{5}\frac{(s+\bar s) - (c+\bar c)}{\sum q+\bar q}\right\},
\ee
where $\langle e^2\rangle = (e_u^2+e_d^2)/2 = 5/18$ is the mean square charge 
of the up and down quarks.
An analysis~\cite{CCFR92a} using the BCDMS and CCFR data yields
$\langle e^2\rangle = (1.00\pm0.03) \cdot 5/18$.
Neutrino data are taken with heavy nuclear targets and must be corrected
for both, nuclear effects and non-isoscalarity, before being compared to 
deuteron data from charged-lepton experiments.
These corrections are assumed to be equal to those in charged-lepton 
scattering where they are well measured.~\cite{E139_94,NMC95a,NMC96c}
In addition to $F_2$ the parity-violating, nonsinglet structure function
$xF_3$, which measures the difference of quark and antiquark
contributions, can be studied in neutrino and antineutrino scattering.

The most precise data for deep-inelastic neutrino scattering come from
the CCFR Collaboration at FNAL. 
The complete $F_2$ and $xF_3$ data sets are shown in Fig.~\ref{fig:ccfr_1}.
For $x\ge0.1$ good agreement of the
corrected CCFR iron $F_2$ data~\cite{CCFR92a,CCFR97a} with the NMC and BCDMS 
deuteron data is found. However, in the small region, $x<0.01$, 
the neutrino data are up to 20~\% larger (Fig.~\ref{fig:ccfr_2}). 
In the first $x$ bins this discrepancy, which decreases systematically with 
increasing $x$, is outside the statistical and  systematic errors.
In this region the strange sea contribution may become important.
The strange-quark content of the nucleon was determined directly~\cite{CCFR95} 
from neutrino-induced charm production, $\nu_\mu s \rightarrow \mu^- c$, and
analysed in terms of the parameter
\be
\kappa=\frac{\int_0^1 xs(x)+x\bar s(x)\,{\rm d}x}
            {\int_0^1 x\bar u(x)+x\bar d(x)\,{\rm d}x}.
\ee
In a next-to-leading order analysis a value of $\kappa=0.477\pm0.051$
was found establishing a suppression of the strange sea.
The shape of the strange quark distribution function, $s(x)$,  is compatible
with that of the light antiquarks.
A possible difference between the strange and antistrange distribution 
functions, 
$s(x) \ne \bar s(x)$, which is expected for an intrinsic strangeness 
component in the nucleon,~\cite{BrP81,BuW92} is beyond the precision 
of the present data. A fit allowing different shapes for $s(x)$ and 
$\bar s(x)$,  with $s(x)/\bar s(x) =A(1-x)^{\Delta\alpha}$, 
yielded $\Delta\alpha=-0.46\pm0.87$.
The strangeness content found in this measurement is not sufficient
to explain the discrepancy between the charged-lepton and neutrino
$F_2$ data at $x\le0.1$.
Possible other sources
for the discrepancy are discussed in Refs.~[\citen{BaG91,BaG93,DoL94}]

\section{QCD analyses and determination of $\alpha_s$}

Perturbative QCD predicts the $Q^2$ evolution of the flavour singlet,
$q^{\rm s}(x,Q^2)=\sum_f q_f(x,Q^2)$, and nonsinglet,
$q^{\rm ns}(x,Q^2)=q_i(x,Q^2)-q_j(x,Q^2)$, quark distribution
functions and of the gluon distribution function, $g(x,Q^2)$.
The quark singlet combination evolves coupled to the gluons
while quark nonsinglet combinations evolve independently.
Typical nonsinglet combinations are the difference of quarks
and antiquarks, $xF_3$, and the difference of up and
down quarks, $F_2^{\rm p}-F_2^{\rm n}$, while $F_2^{\rm d}$ 
is almost a pure singlet combination. 
Apart from a test of QCD the aim of QCD analyses of the 
structure function data is the  determination of the parton 
distribution functions and of the strong coupling constant, 
$\alpha_s$. 
The parton distribution functions are parametrised at a starting
scale $Q^2_0$ and are then evolved to the
$Q^2$ of the data points according to the GLAP equations.

\begin{figure}[t]
   \begin{minipage}[t]{0.49\hsize}
      \begin{center}
         \mbox{\epsfxsize=\hsize\epsfbox[124 519 470 865]{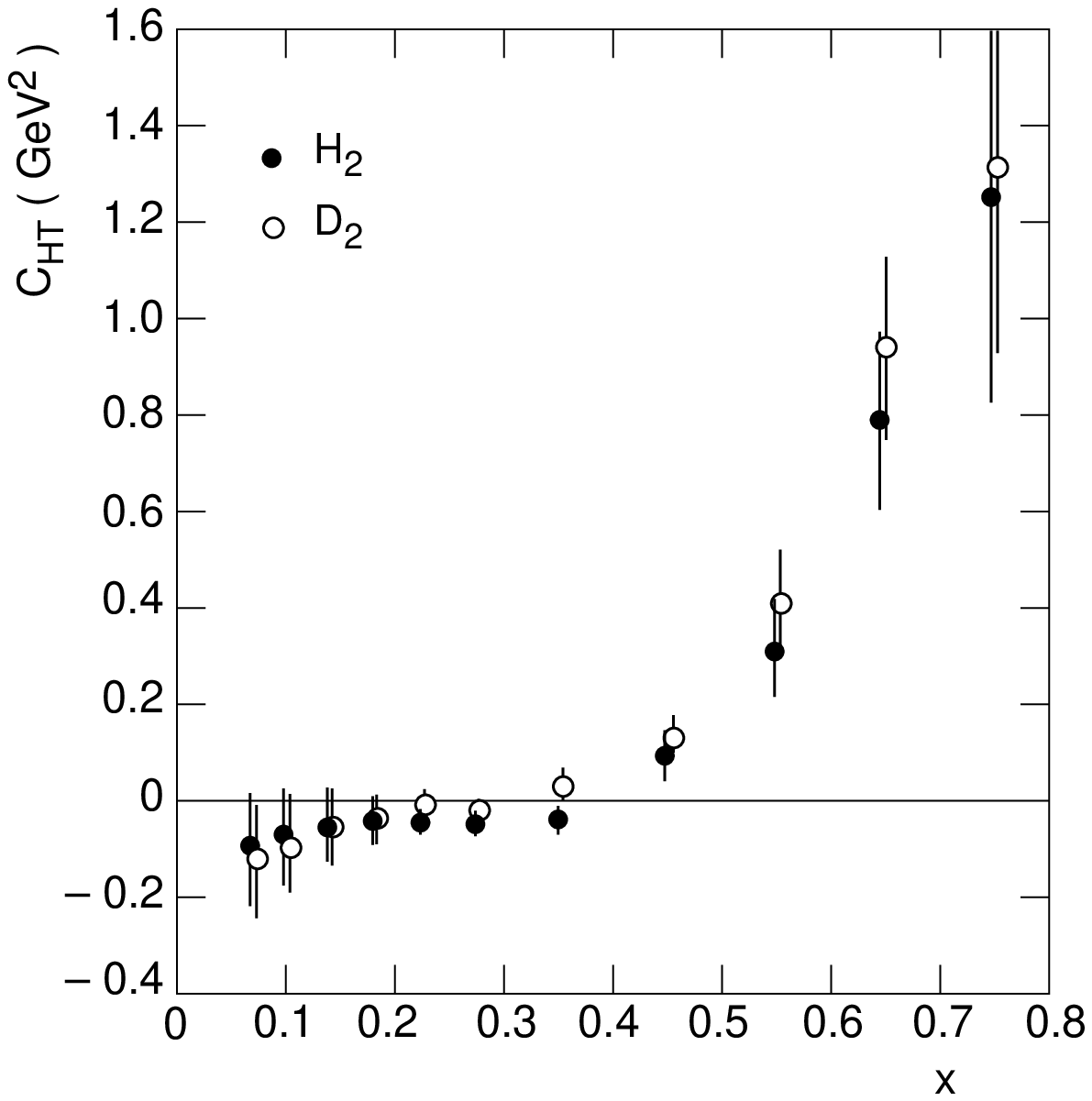}}
         \end{center}
      \caption{Higher-twist coefficients for the proton and the
               deuteron as a function of $x$.~\protect\cite{ViM92}}
      \label{fig:htbcdms}
      \end{minipage}
   \hfill
   \begin{minipage}[t]{0.49\hsize}
      \begin{center}
         \mbox{\epsfxsize=\hsize\epsfbox[5 63 561 610]{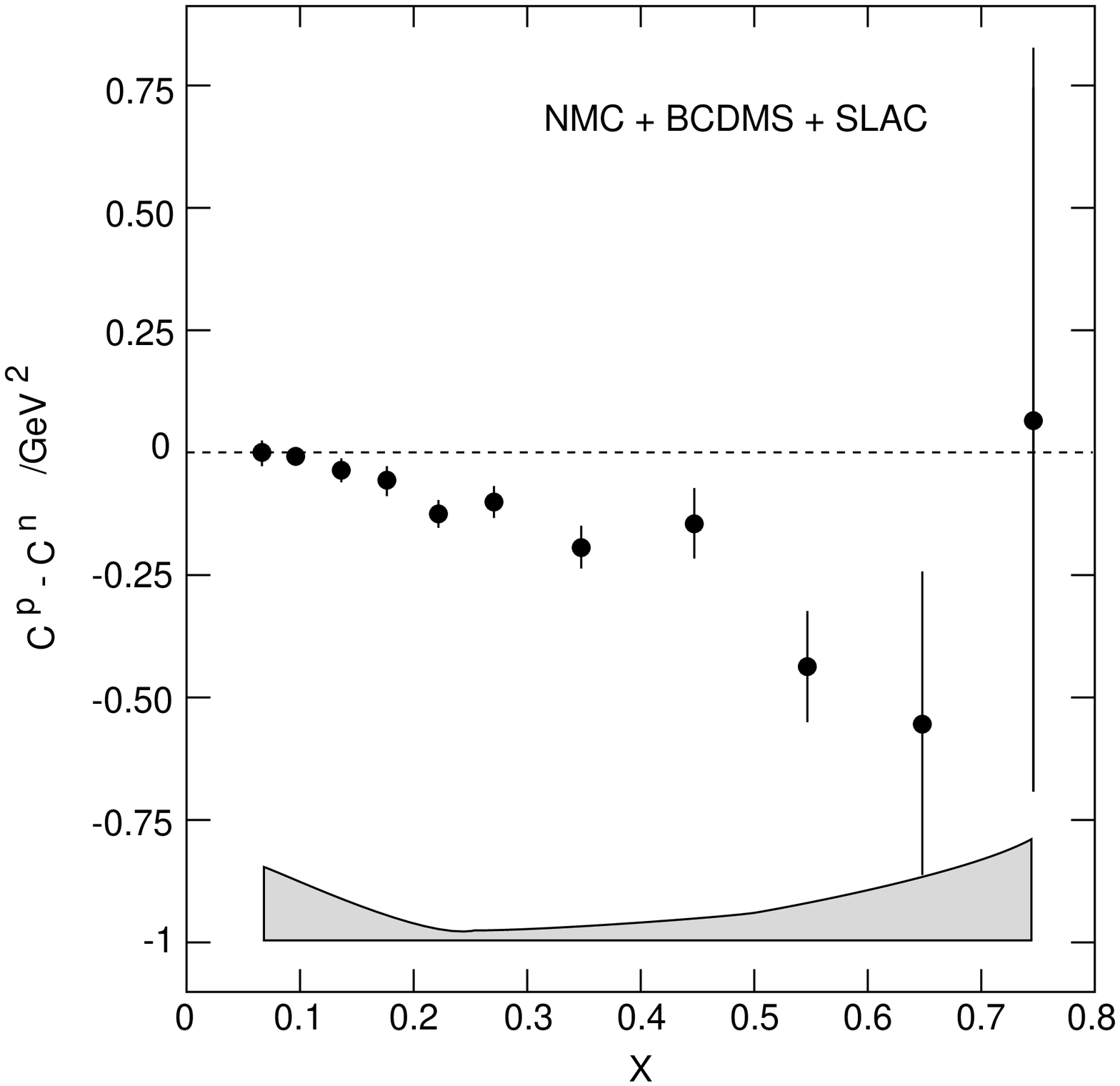}}
        \end{center}
      \caption{The difference of the higher twist coefficients of the
               proton and the deuteron as a function of $x$.~\protect\cite{NMC92b}}
      \label{fig:htnmc}
      \end{minipage}
   \end{figure}

For the BCDMS/SLAC,~\cite{ViM92} the NMC,~\cite{NMC93} and the 
CCFR~\cite{CCFR97a} data next-to-leading order QCD analyses were carried
out by the experimental groups. In the BCDMS and NMC analyses the
proton and deuteron data were fitted simultaneously.
The data included in the BCDMS fit cover $0.07<x<0.85$ with
$Q^2$ in the range 0.5--260~GeV$^2$. 
In addition to the leading-twist contribution and target-mass 
corrections,~\cite{GeP76} combined in $F_2^{LT}$, higher-twist corrections
had to be included to describe the data. They were parametrised
in the form
\be
F_2(x_i,Q^2) = F_2^{LT}(x_i,Q^2)\left\{1+\frac{c_i}{Q^2}\right\},
\ee
where $c_i$ is the higher-twist coefficient for the  $i$\,th $x$ bin.
The fitted coefficients for the proton, $c_i^{\rm p}$, and for the
deuteron, $c_i^{\rm d}$, are shown in Fig.~\ref{fig:htbcdms}.
A calculation of these coefficients using infrared renormalons
is in good agreement with the experimental results.~\cite{DaW96}
The differences of the proton and neutron coefficients, 
$c_i^{\rm p}-c_i^{\rm n}$ were studied with higher precision in a 
combined analysis~\cite{NMC92b}
of $F_2^{\rm n}/F_2^{\rm p}$ data from the NMC, BCDMS, and SLAC
experiments (Fig.~\ref{fig:htnmc}).

\begin{figure}[t]
   \begin{minipage}[t]{0.49\hsize}
      \begin{center}
         \mbox{\epsfxsize=\hsize\epsfbox[26 45 508 508]{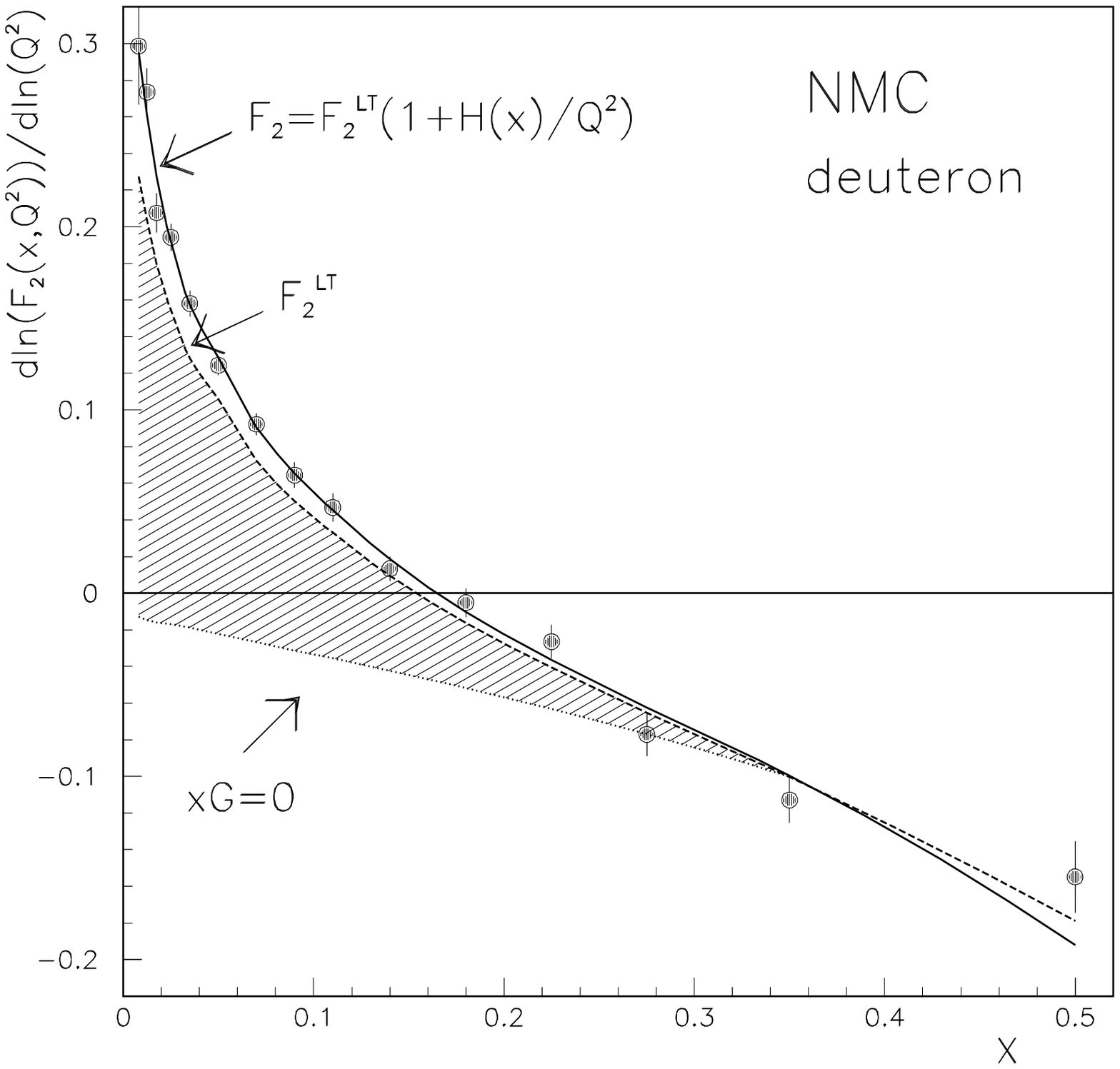}}
         \end{center}
      \caption{The logarithmic slopes
               ${\rm d}\ln F_2/$ ${\rm d}\ln Q^2$
               of the
               NMC deuteron data as a function of $x$.  The solid line shows
               the slopes obtained from the NLO QCD analysis.~\protect\cite{NMC93}
               The dashed line correspond the the QCD prediction without
               higher-twist terms and the dotted line to the $Q^2$ evolution 
               due to quarks only.}
      \label{fig:NMC93_slp}
      \end{minipage}
   \hfill
   \begin{minipage}[t]{0.49\hsize}
      \begin{center}
         \mbox{\epsfxsize=\hsize\epsfbox[26 45 508 508]{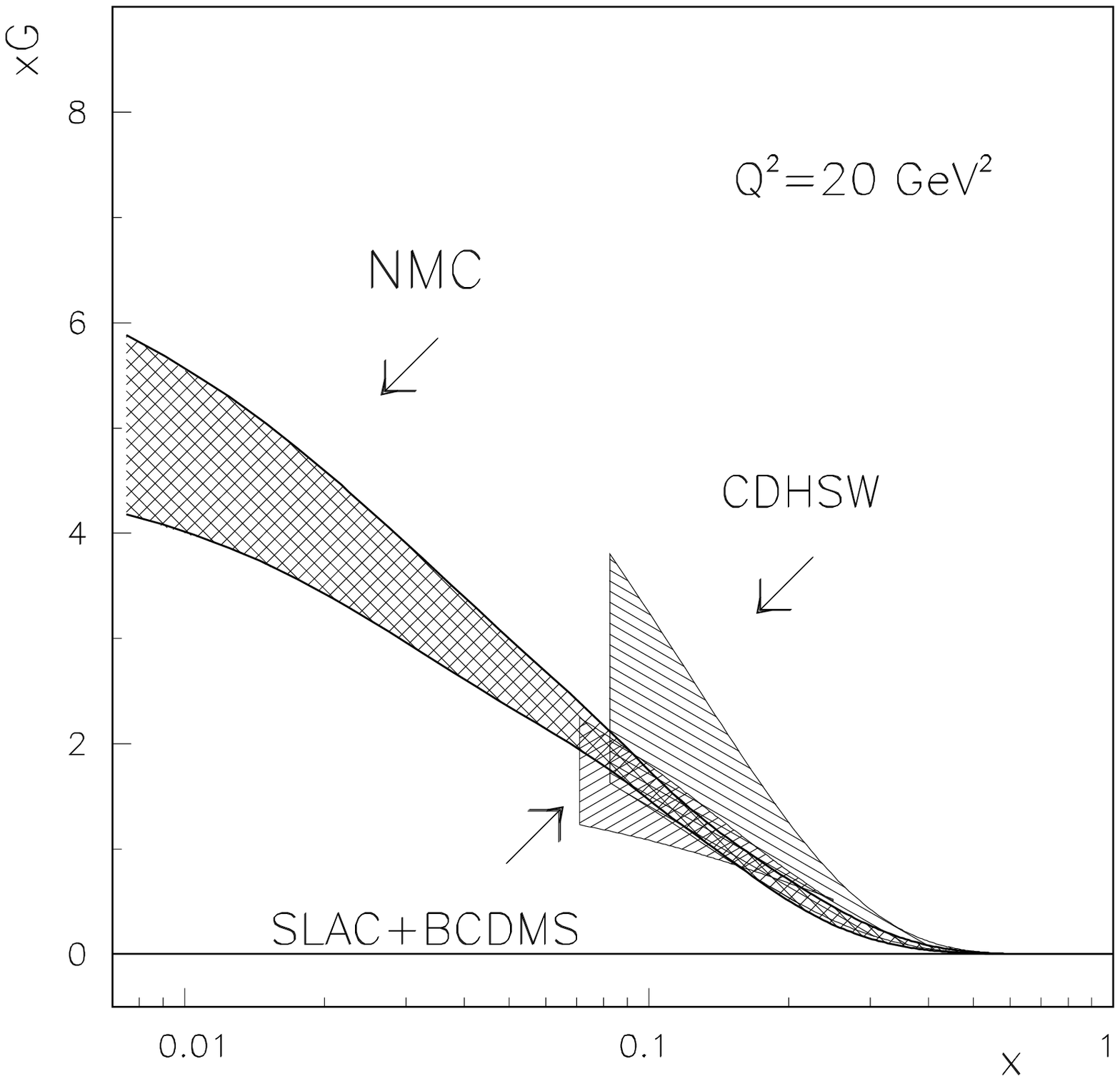}}
        \end{center}
      \caption{The gluon distribution function from the NMC QCD 
                analysis~\protect\cite{NMC93} as function of $x$ at 20~GeV$^2$
                compared to that from the analysis of the BCDMS/SLAC 
                data~\protect\cite{ViM92} and that of a LO analysis of
                CDHSW data.~\protect\cite{CDHSW91a}}
      \label{fig:NMC93_glue}
      \end{minipage}
   \end{figure}

The higher-twist coefficients from the BCDMS analysis were used in the
next-to-leading order QCD analysis~\cite{NMC93} of the NMC data with 
$Q^2>1~\mbox{GeV}^2$.
This analysis was the first to focus on the small $x$ region, 
$0.008<x<0.5$, where the gluon contribution dominates. 
This is demonstrated in Fig.~\ref{fig:NMC93_slp}, where the total QCD 
evolution and that due to quarks as obtained from the QCD fit 
is shown separately.
At $x=0.01$ the gluon distribution function was determined with a 
precision of 20~\% (Fig.~\ref{fig:NMC93_glue}). 
The total momentum fraction carried by quarks and gluons in the 
region covered by the 
data amounts to 0.95 and the quarks carry a momentum fraction of 
$0.55\pm0.02$ at $Q^2_0=7~\mbox{GeV}^2$.

In  the QCD analysis~\cite{CCFR97a} of the CCFR $F_2$ and $xF_3$ structure 
functions, data with $Q^2<5~\mbox{GeV}^2$ or a hadronic final-state energy
of $W^2<10~\mbox{GeV}^2$ were excluded.
This largely removed the small-$x$ region, where the charged-lepton and
the neutrino $F_2$ data differ. The fit is shown as solid line in 
Fig.~\ref{fig:ccfr_1}. 
The value obtained for the strong coupling constant,
$\alpha_s(M_{\rm Z}^2)=0.119\pm0.004$, is considerably larger than
the one found in a previous analysis~\cite{CCFR93b} of the same data,
$\alpha_s(M_{\rm Z}^2)=0.111\pm0.004$. 
The change by two standard deviations is mainly attributed to changes in the 
energy calibration.
The results for $\alpha_s(M_{\rm Z}^2)$ are summarised in Table~\ref{tab:alpha}
together with some typical parameters of the fits. The contributions from
the different error sources were added in quadrature.

\begin{figure}[t]
\mbox{\epsfxsize=\hsize\epsfbox[0 0 480 227]{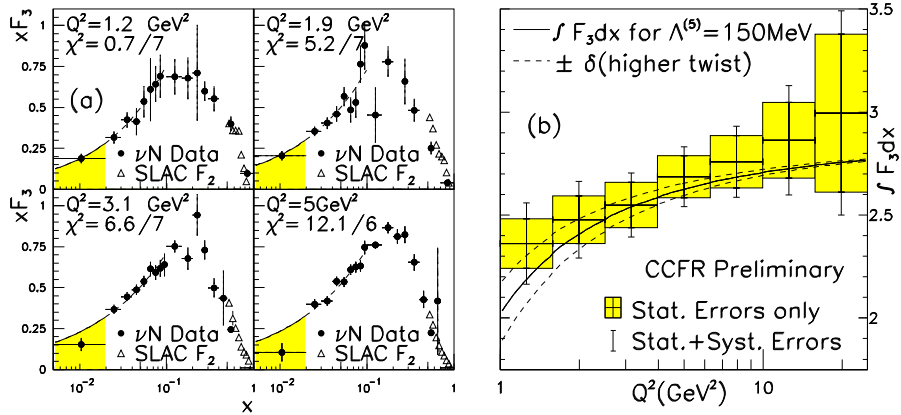}}
\caption{
(a) The structure function $xF_3$ as function of $x$ for four $Q^2$ bins.
(b) GLS sum as function of $Q^2$.~\protect\cite{CCFR95a}}
\label{fig:ccfr_gls}
\end{figure}

\begin{table}
\caption{Results for $\alpha_s$ from NLO QCD analyses of DIS data}
\label{tab:alpha}
\begin{center}
\renewcommand{\arraystretch}{1.1}
\begin{tabular}{|lcrrr@{}r|}
\hline
\hline
Data       & Ref. 
&\multicolumn{1}{c}{ $Q^2_{\rm min}$} 
&\multicolumn{1}{c}{$Q_0^2$}  
&\multicolumn{2}{c|}{$\alpha_s(M_{\rm Z}^2)$} \\
\hline
BCDMS/SLAC &\citen{ViM92}  & 0.5~GeV$^2$ & 20~GeV$^2$ & 0.113&$\pm0.005$ \\
NMC        &\citen{NMC93}  & 1.0~GeV$^2$ &  7~GeV$^2$ & 0.117&$^{+0.011}_{-0.016}$\\
CCFR       &\citen{CCFR97a}& 5.0~GeV$^2$ &  5~GeV$^2$ & 0.119&$\pm0.004$ \\
HERA       &\citen{BaF95c} &             &            & 0.120&$\pm0.010$ \\\hline
CCFR (GLS, &\citen{ChK92}  &                          &3.0~GeV$^2$ & 0.115&$\pm0.006$\\
\multicolumn{1}{|r}{NNLO)}&\citen{CCFR95a}&                          & 1--20~GeV$^2$ & 0.108&$^{+0.007}_{-0.009}$\\\hline
PDG        &\citen{PDG96}  &                          &               & 0.118&$\pm0.003$\\\hline
\hline
\end{tabular}
\end{center}
\end{table}

The $Q^2$ dependence of moments of structure functions is directly
predicted by the operator-product expansion. The GLAP equations
can be obtained by a transformation from moment to coordinate space.
Therefore the study of moments of structure functions is particularly
interesting from the theoretical point of view. However, the extrapolations
to $x=0$ (and $x=1$) and the fact that the sum rules must be evaluated
at a fixed value of $Q^2$ for all values of $x$ introduces additional experimental 
uncertainties.
The Gross--Llewellyn Smith sum rule~\cite{GrL69} (GLS) for the 
parity-violating
structure function $xF_3$ states that the nucleon contains three valence
quarks
\be
S_{GLS}=\frac{1}{2}\int_0^1 \frac{xF_3}{x}\, {\rm d}x = 
3\,\left(1-\frac{\alpha_s}{\pi}-{\cal O}(\alpha_s^2)\right).
\ee
The valence quarks represent a flavour-nonsinglet quark combination 
like the difference of up and down quark polarisations, 
$\Delta u-\Delta d$, in the famous Bjorken polarisation sum 
rule.~\cite{Bjo66}
The QCD corrections for both sum rules are known to order 
${\cal O}(\alpha_s^3)$.~\cite{LaV91} 
The CCFR Collaboration has evaluated the GLS sum
at $Q^2=3~\mbox{GeV}^2$ as
$S_{GLS}=2.50\pm0.018\mbox{(stat.)}\pm0.078\mbox{(syst.)}$.~\cite{CCFR93a}
This yields in next-to-next-to-leading order
$\alpha_s(M_{\rm Z}^2)=0.115\pm0.006$.~\cite{ChK92}
In this analysis the structure functions were evolved from the
$Q^2$ of the measurement to 3~GeV$^2$ using the GLAP equations.
To avoid any $Q^2$ evolution of the data a new analysis
was performed~\cite{CCFR95a} including data from other experiments.
In particular, $F_2$ data from SLAC were used for $x>0.5$, where
the antiquark contribution becomes negligible and thus $xF_3\simeq F_2$.
The GLS sum  was evaluated for seven $Q^2$ bins between 1 and 
20~GeV$^2$ (Fig.~\ref{fig:ccfr_gls}).
The data with $Q^2<7$~GeV$^2$ average to
$\alpha_s(M_{\rm Z}^2)=0.108^{+0.007}_{-0.009}$. This value  differs
considerably from the one found in the re-analysis of scaling violations
in $F_2$ and $xF_3$ and from the present world average 
$\alpha_s(M_{\rm Z}^2)=0.118\pm0.003$.~\cite{PDG96}

\begin{figure}[p]
\begin{center}
\mbox{\epsfxsize=\hsize\epsfbox[20 141 548 685]{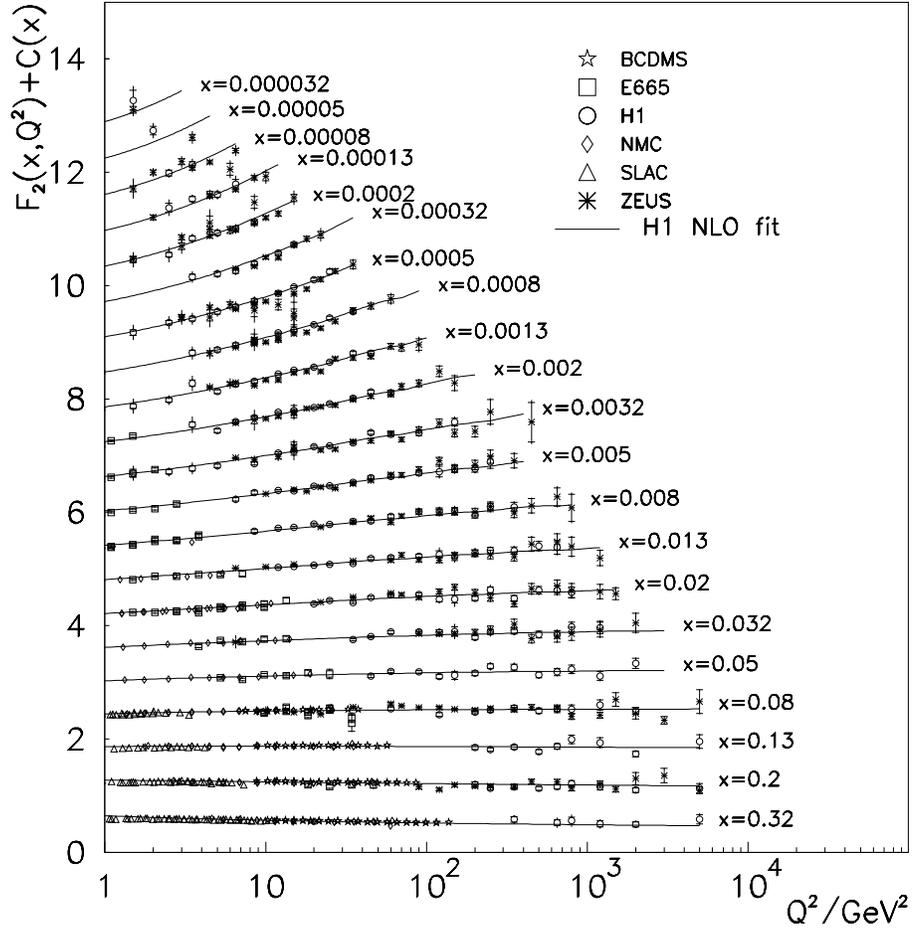}}
\end{center}
\caption{
Hera $F_2$ data~\protect\cite{H1_96a,ZEUS96a} from the 1994 run as function
of $Q^2$ for different values of $x$ together 
with those from fixed-target experiments. Also shown is a NLO QCD 
fit~\protect\cite{H1_96a} to the H1, NMC, and BCDMS data with 
$Q^2>5~\mbox{GeV}^2$. The constants $C(x)$ are defined by 
$C(x)=0.6(i-0.4)$ with the $x$-bin number $i$ and $i=1$ for
$x=0.32$.}
\label{fig:hera96_f2q}
\end{figure}

\begin{figure}[p]
\begin{center}
\mbox{\epsfxsize=\hsize\epsfbox[33 148 548 685]{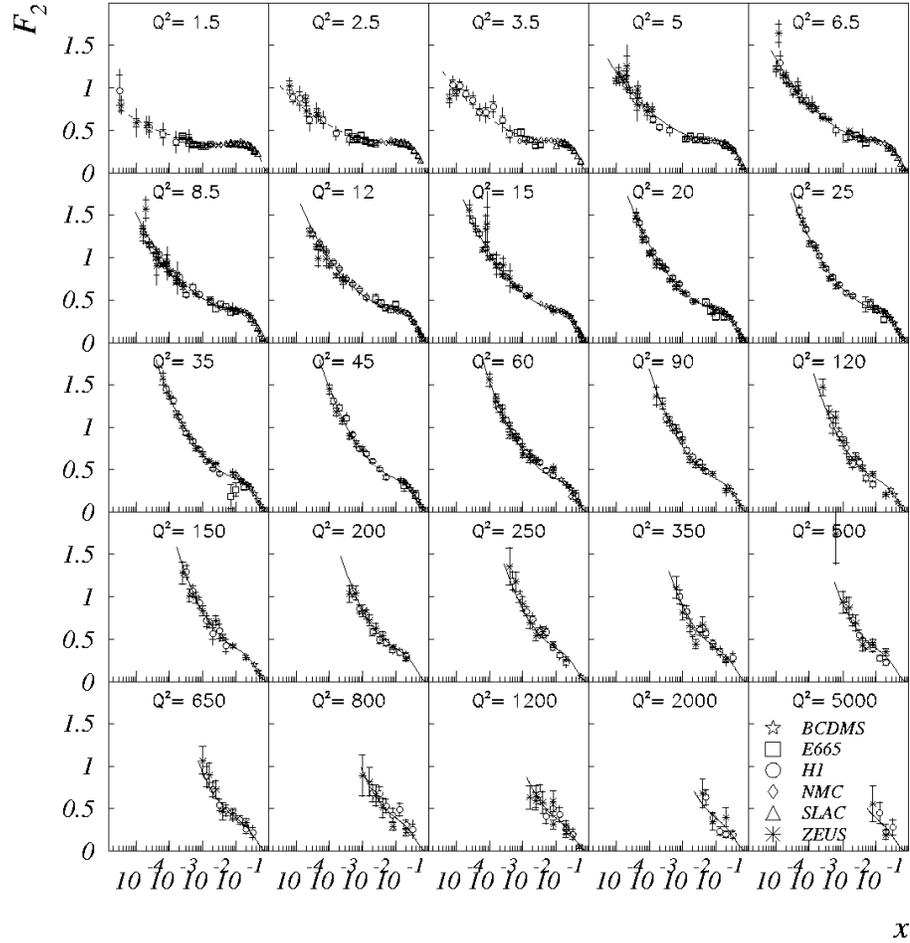}}
\end{center}
\caption{
Hera $F_2$ data~\protect\cite{H1_96a,ZEUS96a} from the 1994 run as function
of $x$ for different values of $Q^2$ together 
with data from fixed-target experiments. Also shown is a NLO QCD 
fit~\protect\cite{H1_96a} to the H1, NMC, and BCDMS data with 
$Q^2>5~\mbox{GeV}^2$.}
\label{fig:hera96_f2x}
\end{figure}

\begin{figure}[t]
\begin{center}
\mbox{\epsfxsize=\hsize\epsfbox[32 247 536 595]{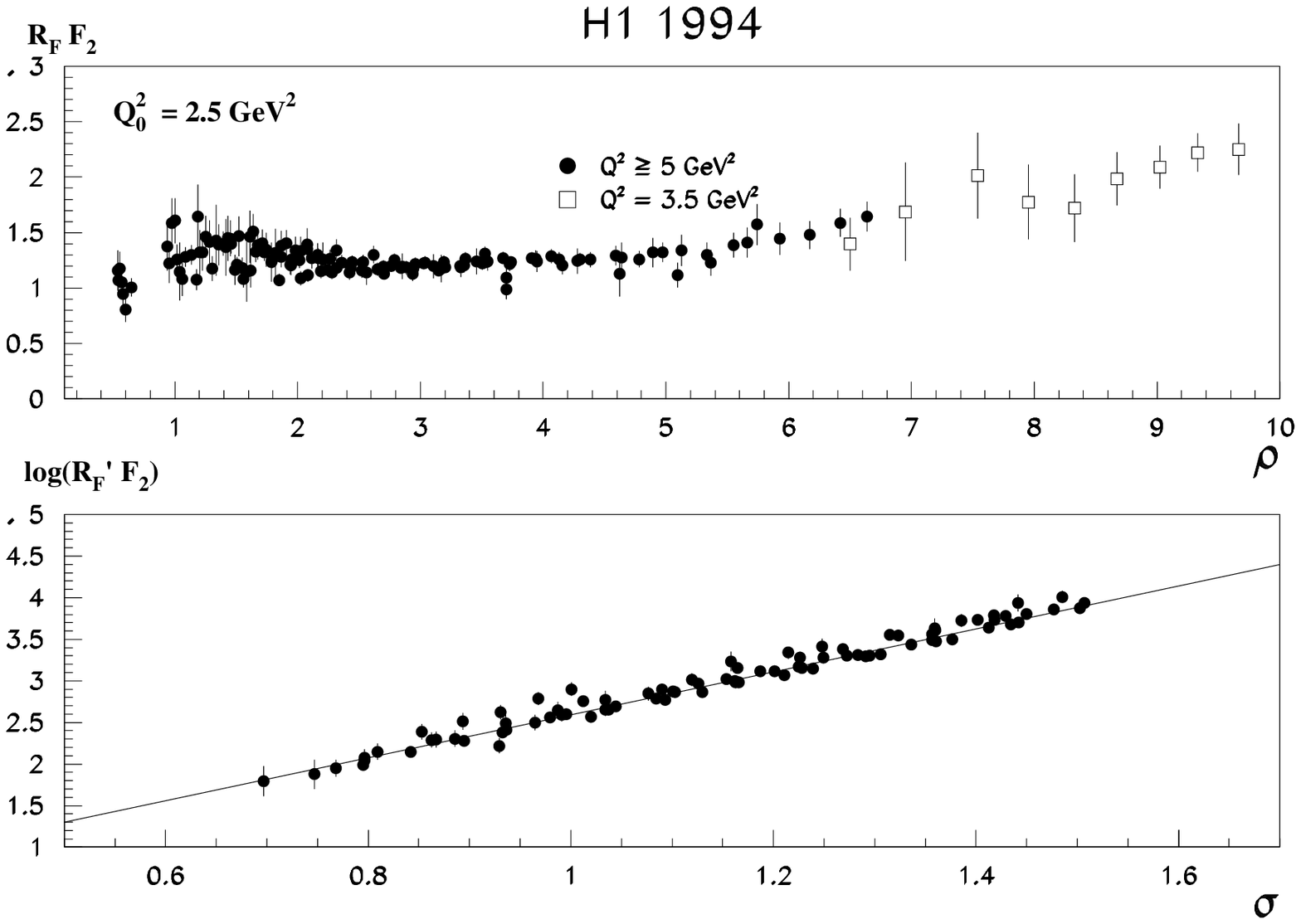}}
\end{center}
\caption{
The rescaled structure function $R_F F_2$ as a function of $\rho$ (top)
and $\ln R_F' F_2$ as a function of $\sigma$~\protect\cite{H1_96a} (bottom).}
\label{fig:hera96_2log}
\end{figure}

\section{HERA data for $F_2^{\rm p}$}
The HERA experiments H1 and ZEUS~\cite{Zeuner} have extended the kinematic
range of $F_2^{\rm p}$ measurements down to $x\simeq 10^{-5}$ and to 
$Q^2>5000$~GeV$^2$ for $x>0.1$. 
The  HERA $F_2$ data~\cite{H1_96a,ZEUS96a} taken in 1994 agree 
fairly with the high
$Q^2$ extrapolation of the fixed-target data as shown in 
Figs.~\ref{fig:hera96_f2q}, \ref{fig:hera96_f2x}, and \ref{fig:NMC_HERA}.
In the small-$x$ region
a steep rise of $F_2\sim x^{-\lambda}$ is observed for decreasing $x$.
The parameter $\lambda$ increases with $Q^2$ from about 0.2 at 
10~GeV$^2$ to 0.3 at 100~GeV$^2$ and possibly 
to 0.5 at 1000~GeV$^2$.~\cite{H1_96a}
The rise of $F_2$ towards small $x$ continues down to 
$Q^2\simeq1.5~\mbox{GeV}^2$. Thus Regge-inspired models underestimate
$F_2$ in this region. However, preliminary results from ZEUS~\cite{ZEUS96b}
using the beam-pipe calorimeter and reaching down to $Q^2=0.16~\mbox{GeV}^2$
and $x\simeq3\times10^{-6}$ agree with the
Donnachie--Landshoff model~\cite{DoL92} below $Q^2=1~\mbox{GeV}^2$.

The asymptotic behaviour of $F_2$ in the small-$x$--large-$Q^2$ range
is generated by QCD dynamics~\cite{DeR74,BaF94a} and can be described by the 
two variables,
\be 
\sigma=\sqrt{\xi\zeta}\hskip 0.5cm \mbox{and} \hskip 0.5cm
\rho = \sqrt{\xi/\zeta},
\ee
with $\xi=\ln(x_0/x)$ and $\zeta=\ln\ln (Q^2/Q_0^2)$. In the HERA region
$\ln F_2$ should grow linearly with $\sigma$ and be independent
of $\rho$. 
The $\sigma$ slope can be calculated and one obtains 
${\rm d}\ln F_2/{\rm d}\sigma =2\gamma=2.5$ for five active flavours.
With $R_F=R_F'\exp(-2\gamma)$
the quantity $R_F F_2$ becomes independent of both, $\sigma$ and $\rho$.
Here $R_F'$ accounts for the finite size of $\sigma$ and $\rho$. 
The HERA data impressively confirm this asymptotic behaviour as shown
in Fig.~\ref{fig:hera96_2log}. 
The $\sigma$ slope of $\ln(R_F'F_2)$ fitted to the
H1 data~\cite{H1_96a} is $2.57\pm0.08$ in good agreement with the prediction.

The strong correlation of the gluon distribution function with $\alpha_s$ 
at small $x$ makes it difficult to determine $\alpha_s$ in a standard QCD 
analysis from the HERA $F_2$ data alone. However,
the growth of $F_2$ in the small-$x$ and large-$Q^2$ region is directly
related to the value of $\alpha_s$.~\cite{BaF95c}
An analysis of the HERA data taken in 1993 yielded a rather high value 
compared to other DIS results,
$\alpha_s(M_{\rm Z}^2)=0.120\pm0.005\mbox{(exp.)}\pm0.009\mbox{(theory)}$.
A preliminary result from the 1994 data yields an even higher value 
of $0.122\pm0.004\mbox{(exp.)}$.~\cite{BaF96a}

\begin{figure}[t]
   \begin{minipage}[t]{0.49\hsize}
      \begin{center}
         \mbox{\epsfxsize=\hsize\epsfbox[17 151 547 691]{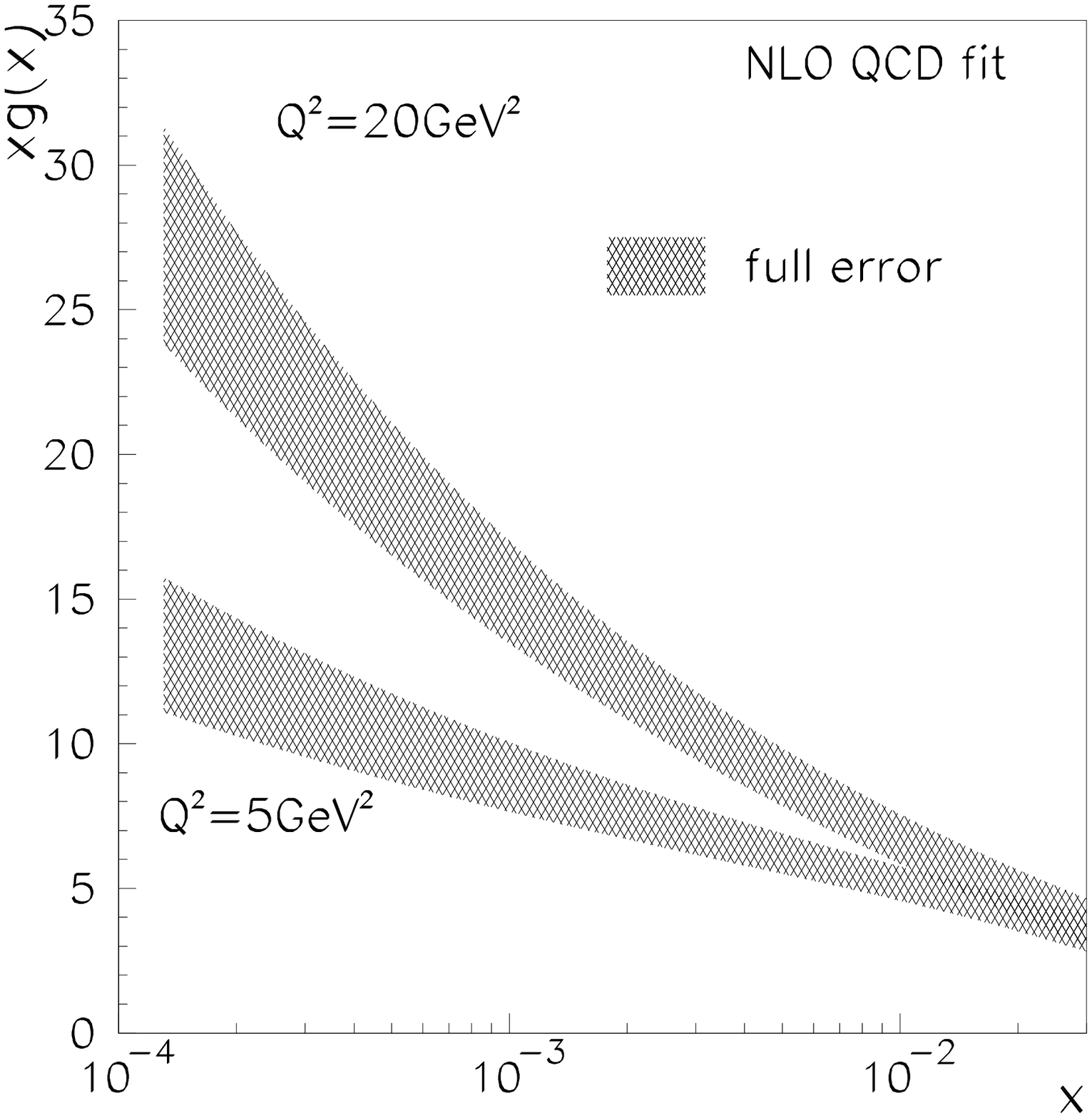}}
         \end{center}
      \caption{The NLO $\overline{\rm MS}$ gluon distribution function, $xg(x)$,
               at $Q^2=5$ and 20~GeV$^2$ as a function of $x$ 
               from a QCD analysis of the H1 
               data.~\protect\cite{H1_96a}}
      \label{fig:h1_96_glue}
      \end{minipage}
   \hfill
   \begin{minipage}[t]{0.49\hsize}
      \begin{center}
         \mbox{\epsfxsize=\hsize\epsfbox[14 5 538 510]{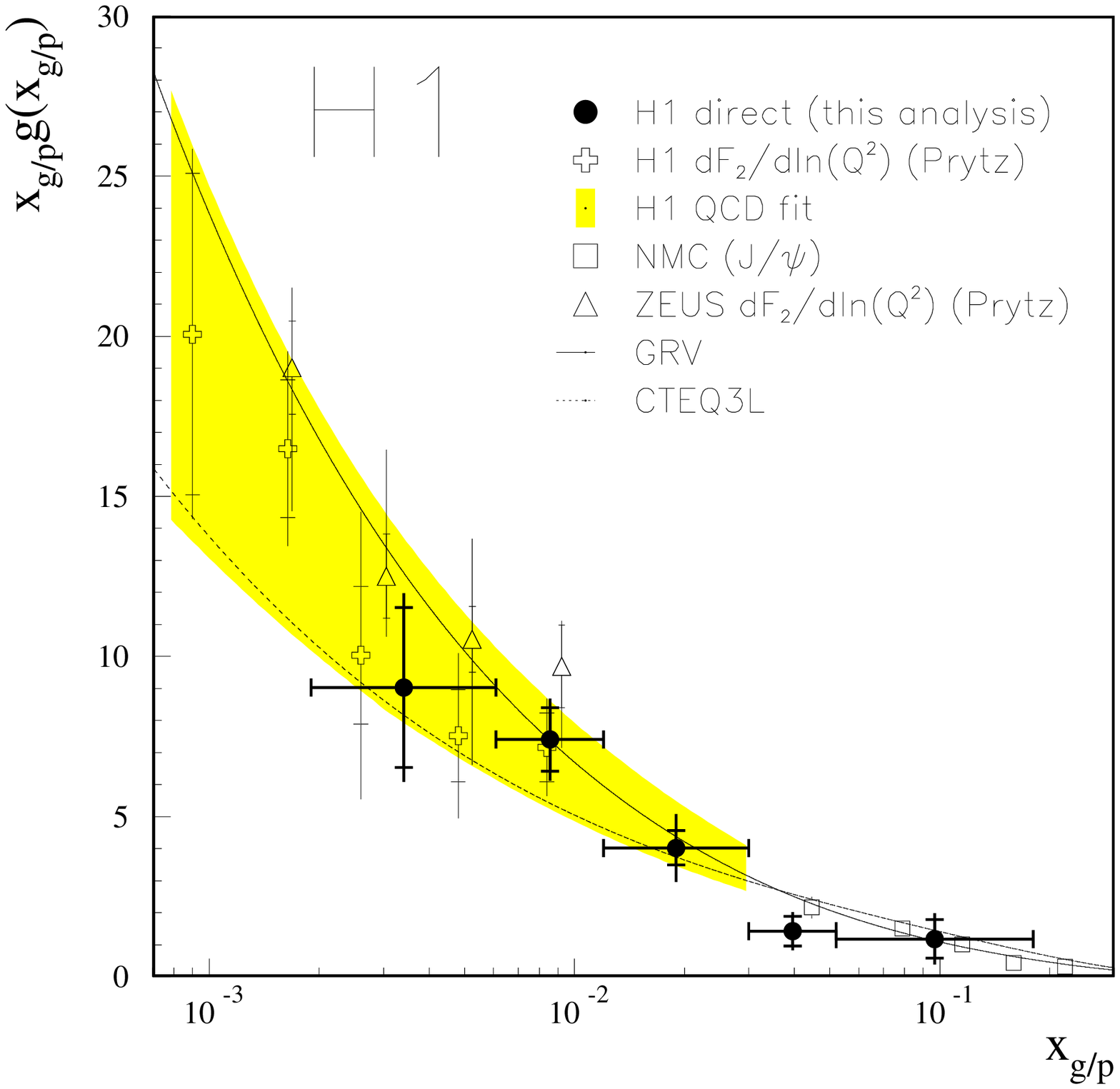}}
        \end{center}
      \caption{The leading-order gluon distribution function, $xg(x)$, at
               $Q^2=30$~GeV$^2$ from (2+1)-jet 
                data~\protect\cite{H1_95a} and from
               J/$\psi$ production.~\protect\cite{NMC91a}}
      \label{fig:h1_93_glue}
      \end{minipage}
   \end{figure}
The gluon distribution function determined in a next-to-leading
order QCD analysis in the $\overline{\rm MS}$ scheme from the H1 
data~\cite{H1_96a} is shown in Fig.~\ref{fig:h1_96_glue}. New parton 
distribution functions from global fits including the new HERA data 
are available from the MRS~\cite{MaR96a} and CTEQ~\cite{LaH97} groups.
A direct measurement of $g(x)$ can be performed using the (2+1)-jet
production data. From the photon-gluon fusion process, 
$\gamma^\star g\rightarrow q\bar q$, two quark jets emerge.
The target jet disappears in the beam pipe.
There is an about 30~\% contribution from the QCD Compton 
process, $\gamma^\star q \rightarrow q'+g$.
The (2+1)-jet production is via the photon-gluon fusion process directly
linked to the gluon distribution in the proton. Leading-order results 
from the 1993 (2+1)-jet data from H1~\cite{H1_95a} are shown in 
Fig.~\ref{fig:h1_93_glue}
together with NMC results from J/$\psi$ production.~\cite{NMC91a}
They are in good agreement with other determinations of $xg(x)$.

\section{Conclusions}
The fixed-target structure function data are in good agreement. A small
discrepancy in the small-$x$ region persists between the muon and neutrino
$F_2$ data. Presently, it is unclear whether this has a physics origin or 
is a reflection of experimental difficulties.
This question is partly related to the r\^ole
of strangeness in the nucleon and maybe also to the longitudinal structure
function, $F_L$. For $F_L$ considerable progress was made in the small-$x$
region. However, it is still much less well known than $F_2$. An
accurate measurement at HERA, requiring a luminosity upgrade, is highly 
desirable.
The theoretical understanding of higher-twist corrections, in particular
of their $x$ dependence, has made considerable progress over the last 
years and the calculations agree well with the data.
This is an essential ingredient for the determination of $\alpha_s$ from 
sum rules, e.g.\ the Gross-Llewellyn Smith sum rule.
While up to now most determinations of $\alpha_s$ from scaling violations
in structure functions yielded consistently smaller results than obtained
from LEP, the situation is less clear after the re-analysis of the CCFR
structure functions.
A determination of $\alpha_s$ from the small-$x$--high-$Q^2$ 
region of the HERA data yields also a rather large result.
The rise of $F_2$ with decreasing $x$ and increasing $Q^2$ has been mapped
out at HERA with a considerable precision. Thus a good determination
of the gluon distributions at small $x$ became possible.
 
Deep inelastic scattering continues to be one of the most important
testing grounds of perturbative -- and to the extent our present understanding
allows -- also of nonperturbative QCD.

\section*{Acknowledgements}
I thank my colleagues in the New Muon and Spin Muon 
Collaborations for many helpful discussions and B.~Frois and 
V.~W.~Hughes for organising this stimulating Workshop in Erice and for
their patience awaiting this manu\-script.

\end{document}